%% file: main.tex
\documentclass[letterpaper,twocolumn,10pt]{article}
\usepackage{usenix2019_v3,epsfig,endnotes}
\usepackage{times}

\usepackage{tikz}
\usepackage{booktabs} 
\usepackage{multirow}
\usepackage{tabularx}
\usepackage{url}
\usepackage{paralist} 
\usepackage{amsthm}
\usepackage{color}
\usepackage{comment}
\usepackage[cmex10]{amsmath}
\usepackage{listings}
\usepackage{graphicx}
\usepackage{xspace}
\PassOptionsToPackage{hyphens}{url}\usepackage{hyperref}
\usepackage[singlelinecheck=on,font=footnotesize,labelfont=bf]{caption}
\usepackage{makecell}
\usepackage[subtle]{savetrees}  
\usepackage{soul}
\usepackage{float}
\usepackage{footnote}
\usepackage{fancyvrb}
\usepackage{subcaption} 
\usepackage{cite}
\usepackage{titlesec}
\usepackage{xcolor}
\usepackage[titletoc,toc,title]{appendix}
\titlespacing*{\section}
{0pt}{3ex}{2ex}
\titlespacing*{\subsection}
{0pt}{2ex}{1ex}

\renewcommand\appendix{\par
  \setcounter{section}{0}
  \setcounter{subsection}{0}
  \setcounter{figure}{0}
  \setcounter{table}{0}
}

\newcommand{\point}[1]{\vspace{0.8ex}\par\noindent{\bf #1:}}
\newcommand{\emphasize}[1]{\vspace{0.8ex}\par\noindent{\bf #1}}
\newcommand{\etc}{\textit{etc.}\xspace}
\newcommand{\eg}{\textit{e.g.,}\xspace}
\newcommand{\ie}{\textit{i.e.,}\xspace}
\newcommand*{\rom}[1]{\textit{\expandafter\romannumeral #1}}

\newcommand{\mlappnum}{1,468\xspace}
\newcommand{\mlapprate}{3.14\%\xspace}
\newcommand{\insight}[1]{\textit{\ul{ #1}}}

\newcommand{\qone}{How widely is model protection used in apps?\xspace}
\newcommand{\qtwo}{How robust are existing model protection techniques?\xspace}
\newcommand{\qthree}{What impacts can (stolen) models incur?\xspace}

\newcommand{\Qone}{How Widely Is Model Protection Used in Apps?\xspace}
\newcommand{\Qtwo}{How Robust Are Existing Model Protection Techniques?\xspace}
\newcommand{\Qthree}{What Impacts can (Stolen) Models Incur?\xspace}

\newcommand{\toolmxr}{ModelXRay\xspace}
\newcommand{\toolmxt}{ModelXtractor\xspace}
\newcommand{\anony}[2]{#2}

\usepackage{filecontents}
\renewenvironment{itemize}[1]{\begin{compactitem}#1}{\end{compactitem}}

\theoremstyle{definition}

\def\zc{0}

\ifx\zc\undefined
	
\else
	
\fi

\definecolor{mygreen}{rgb}{0,0.6,0}
\definecolor{mygray}{rgb}{0.5,0.5,0.5}
\definecolor{mymauve}{rgb}{0.58,0,0.82}

\begin{document}
\twocolumn

\date{}
\title{\Large \bf Mind Your Weight(s): A Large-scale Study on Insufficient \\
 Machine Learning Model Protection in Mobile Apps} 

\author{
{\rm Zhichuang Sun}\\
Northeastern University
\and
{\rm Ruimin Sun}\\
Northeastern University
 \and
{\rm Long Lu}\\
Northeastern University
 \and
{\rm  Alan Mislove }\\
Northeastern University
}

\maketitle
\thispagestyle{empty}

\input{abstract}

\input{introduction}
\input{background}
\input{overview}
\input{static_new}

\input{dynamic}

\input{impact}

\input{countermeasures}

\input{discussion}

\input{relatedwork_1}

\input{conclusion}
\input{ack}

 \newpage
\bibliographystyle{plain}
\bibliography{bibliography}

\input{appendix}

\end{document}

%% file: abstract.tex
\begin{abstract}

On-device machine learning (ML) is quickly gaining popularity among mobile apps.
It allows offline model inference while preserving user privacy. However, ML
models, considered as core intellectual properties of model owners, are now
stored on billions of untrusted devices and subject to potential  thefts. Leaked
models can cause both severe financial loss and security consequences. 

This paper presents the first empirical study of ML model protection on mobile
devices. Our study aims to answer three open questions with quantitative
evidence: \qone \qtwo \qthree 
%
To that end, we built a simple app analysis pipeline and
analyzed 46,753 popular apps collected from the US and Chinese app markets. We
identified 1,468 ML apps spanning all popular app categories. We found that,
alarmingly, 41\% of ML apps do not protect their models at all, which can be
trivially stolen from app packages. 
Even for those apps that use model protection or encryption, we were able to
extract the models from 66\% of them via unsophisticated dynamic analysis
techniques. The extracted models are mostly commercial products and used for
face recognition, liveness detection, ID/bank card recognition, and malware
detection. 
%
We quantitatively estimated the potential financial and security impact of a leaked model,
which can amount to millions of dollars for different stakeholders. 

Our study reveals that on-device models are currently at high risk of being
leaked; attackers are highly motivated to steal such models. Drawn from our
large-scale study, we report our insights into this emerging security problem
and discuss the technical challenges, hoping to inspire future research on
robust and practical model protection for mobile devices.

\end{abstract}

%% file: introduction.tex
\section{Introduction} \label{intro}
Mobile app developers have been quickly adopting {\em on-device machine
learning} (ML) techniques to provide artificial intelligence (AI) features, such
as facial recognition, augmented/virtual reality, image processing, voice
assistant, etc. This trend is now boosted by new AI chips available in the
latest smartphones \cite{mobileai}, such as Apple's Bionic neural engine,
Huawei’s neural processing unit, and Qualcomm’s AI-optimized SoCs. 

Compared to performing ML tasks in the cloud, on-device ML
(mostly model inference) offers unique benefits desirable for mobile users as
well as app developers. For example, it avoids sending (private) user data to
the cloud and does not require network connection.
For app developers or ML solution providers, on-device ML greatly reduces the
computation load on their servers. 

On-device ML inference inevitably stores ML models locally on user devices,
which however creates a new security challenge. Commercial ML models used in
apps are often part of the core intellectual property (IP) of vendors. Such
models may fall victim to theft or abuse, if not sufficiently protected. In
fact, on-device ML makes model protection much more challenging than server-side
ML because models are now stored on user devices, which are fundamentally
untrustworthy and may leak models to curious or malicious parties.


The consequences of model leakage are quite severe. First, with a leaked model
goes away the R\&D investment of the model owner, which
often includes human, data, and computing costs. Second, when a proprietary
model is obtained by unethical competitors, the model owner loses the
competitive edge or pricing advantage for its products.
Third, a leaked model facilitates malicious actors to find
adversarial inputs to bypass or confuse the ML systems, which
can lead to not only reputation damages to the vendor but also critical failures
in their products (\eg fingerprint recognition bypass). 


This paper presents the first large-scale study of ML model protection and
theft on mobile devices. 
Our study aims to shed light on the less understood risks and costs of model
leakage/theft in the context of on-device ML. 
We present our study that answers the following questions with ample empirical evidence and observations.  

\begin{itemize}
    \item {\bf Q1: \qone}
    \item {\bf Q2: \qtwo}
    \item {\bf Q3: \qthree}
\end{itemize}




To answer these questions, we collected 46,753 trending Android apps from the US and the Chinese app markets. To answer Q1, 
we built a simple and automatic pipeline to first identify the ML models and SDK/frameworks used in an app, and then detect if the ML models are encrypted. 
Among all the collected apps, we found 1,468 apps that use on-device ML, and 602 (41\%) of them do not protect their ML models at all (\ie models are stored in plaintext form on devices). Most
of these apps have high installation counts (greater than 10M) and span
the top-ten app categories, which
underlines the limited awareness of model thefts and the need for model
protection among app developers. 


To answer Q2, for the encrypted models, we dynamically run the corresponding apps and built an automatic pipeline to identify and extract the decrypted ML models from memory. This pipeline represents an unsophisticated model theft attack that an adversary can realistically launch on her own device. 
We found that the same protected models can be reused/shared by multiple apps, and a set of 18 unique models extracted from our dynamic analysis can affect 347 apps (43\% of all the apps with protected models).
These apps cover a wide range of ML frameworks, including
TensorFlow, TFLite, Caffe, SenseTime, Baidu, Face++, etc. They use ML for
various purposes, including face tracking, liveness detection, OCR, ID card and
bank card recognition, photo processing, and even malware detection.


We also observed some interesting cases where a few model owners spent extra effort on
protecting their models, such as encrypting both code and model files, encrypting
model files multiple times, or encrypting feature vectors. Despite the efforts,
these models can be successfully extracted in memory in plaintext. These cases
indicate that model owners or app developers start realizing the risk of model
thefts but no standard and robust model protection technique exists, 
which echos the urgent need for research into on-device model protection.

Finally, to answer Q3, we present an analysis on the financial and security impact of model
leakage on both the attackers and the model vendors. We
identify three major sources of impact: the research and development investment on the
ML models, the financial loss due to competition, and the security impact due to 
model evasion. We found that the
potential financial loss can be as high as millions of dollars, depending on the
app revenue and the actual cost of the models. The security impact includes
bypassing the model-based access control, which may result in reputation damage or even
product failure.

By performing the large-scale study and finding answers to the three questions,
we intend to raise the awareness of the model leak/theft risks, which apps using
on-device ML are facing even if models are encrypted. Our study shows that the
risks are realistic due to absent or weak protection of on-device models. It
also shows that attackers are not only technically able to, but also highly
motivated to steal or abuse on-device ML models. We share our insights and call
for future research to address this emerging security problem.

In summary, the contributions of our research are:

\begin{itemize}
     \item We apply our analysis pipeline on 46,753 Android apps collected from
     US and Chinese app markets. We found that among the 1,468 apps using
     on-device ML, 41\% do not have any protection on their ML models. For those
     do, 66\% of them still leak their models to an unsophisticated
     runtime attack. 
     

    \item We provide a quantified estimate on the financial and security impact of model
    leakage based on case studies. We show that attackers with stolen models can save as high as millions of dollars,
    while vendors can encounter pricing disadvantage and falling market share. 
    Further model evasion may cause illegal access to private information of end users.

    \item Our work calls for research on robust protection mechanisms for ML
    models on mobile devices.  We share our insights gained during the study to
    inform and assist future work on this topic. 
\end{itemize}

The rest of the paper is organized as follows. Section \ref{sec:background}
introduces the background knowledge about on-device ML. Section
\ref{sec:overview} presents an overview of our analysis pipeline. Sections
\ref{sec:static}, \ref{sec:dynamic}, and \ref{sec:financial} answers the
questions Q1, Q2, and Q3, respectively. Section \ref{sec:counter} summarizes the
current model protection practices and their effectiveness. Section \ref{sec:discuss} 
discusses the research insights and
the limitations of our analysis. Section \ref{sec:related} surveys the related
work and Section \ref{sec:conclusion} concludes the paper.

%% file: background.tex
\section{Background} \label{sec:background}
\point{The Trend of On-device Machine Learning} 
Currently, there are two ways for mobile apps to use ML: cloud-based and
on-device. In cloud-based ML, apps send requests to a cloud server, where the ML
inference is performed, and then retrieve the results. The drawbacks include
requiring constant network connections, unsuitable for real-time ML tasks (e.g.,
live object detection), and needing raw user data uploaded to the server.
Recently, on-device ML inference is quickly gaining popularity thanks to the
availability of hardware accelerators on mobile devices and the the ML
frameworks optimized for mobile apps. On-device ML avoids the aforementioned
drawbacks of cloud-based ML. It works without network connections, performs well
in real-time tasks, and seldom needs to send (private) user data off the device.
However, with ML inference tasks and ML models moved from cloud to user devices,
on-device ML raises a new security challenge to model owners and ML service
providers: how to protect the valuable and proprietary ML models now stored and
used on user devices that cannot be trusted.

\point{The Delivery and Protection of On-device Models } 

Typically, on-device ML models are trained by app developers or ML service
providers on servers with rich computing resources (e.g., GPU clusters and
large storage servers). Trained models are shipped with app installation
packages. A model can also be downloaded separately after app installation to
reduce the app package size. Model inference is performed by apps on user
devices, which relies on model files and ML frameworks (or SDKs). To protect
on-device models, some developers encrypt/obfuscate them, or compile them into
app code and ship them as stripped binaries\cite{ConvertingCode,Nihui}. However,
such techniques only make it difficult to reverse a model, rather than strictly
preventing a model from being stolen or reused. 

\point{On-device Machine Learning Frameworks} 
There are tens of popular ML frameworks, such as Google TensorFlow and
TensorFlow Lite \cite{TensorFlow}, Facebook PyTorch and Caffe2
\cite{Caffe2Framework.}, Tencent NCNN \cite{Nihui}, and Apple Core ML
\cite{CoreDocumentation}. Among these frameworks, TensorFlow Lite, Caffe2, NCNN
and Core ML are particularly optimized for mobile apps. 
Different frameworks use different file formats for storing ML models on devices,
including ProtoBuf (.pb, .pbtxt), FlatBuffer (.tflite), MessagePack (.model),
pickle (.pkl), Thrift (.thrift), etc. To mitigate model reverse engineering and
leakage, some companies developed customized or proprietary model formats
\cite{Xu2019,SleeThrift:Implementation}.

\point{On-device Machine Learning Solution Providers}
For cost efficiency and service quality, app developers often use third-party ML
solutions, rather than training their own models or maintaining in-house ML
development teams. The popular providers of ML solutions and services include
Face++ \cite{Face++Services} and SenseTime \cite{SenseTimest}, which sell
offline SDKs (including on-device models) that offer facial recognition, voice
recognition, liveness detection, image processing, Optical Character Recognition
(OCR), and other ML functionalities. By purchasing a license, app developers
can include such SDKs in their apps and use the ML functionalities as
black-boxes. ML solution providers are more motivated to protect their models
because model leakage may severely damage their business \cite{SenseTimest}.



%% file: overview.tex
\section{Analysis Overview} \label{sec:overview}

\begin{figure}
\centering
    \includegraphics[width=0.85\columnwidth]{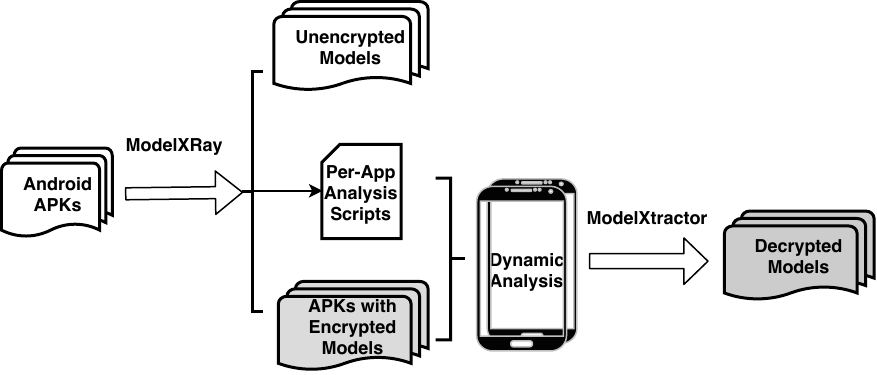}
     \caption{Overview of Static-Dynamic App Analysis Pipeline}
     \label{fig:overview}
\end{figure}

On-device ML is quickly being adopted by apps, while its security
implications on model/app owners remain largely unknown. Especially, the threats
of model thefts and possible ways to protect models have not been sufficiently
studied. This paper aims to shed light on this issue by conducting a large-scale study
and providing quantified answers to three questions: \qone (\S\ref{sec:static})
%
 \qtwo (\S\ref{sec:dynamic})  \qthree (\S\ref{sec:financial})

To answer these questions, we built a static-dynamic app analysis pipeline. We
note that this pipeline and the analysis techniques are kept simple
intentionally and are not part of the research contributions of this work. The
goal of our study is to understand how easy or realistic it is to leak or steal
ML models from mobile apps, rather than demonstrating novel or sophisticated app
analysis and reverse-engineering techniques. Our analysis pipeline represents
what a knowledgeable yet not extremely skilled attacker can already achieve when
trying to steal ML models from existing apps. Therefore, our analysis result
gives the lower bound of (or a conservative estimate on) how severe the model
leak problem currently is.

The workflow of our analysis is depicted in Figure\ref{fig:overview}. Apps first
go through the static analyzer, \toolmxr, which detects the use of on-device ML
and examines the model protection, if any, adopted by the app.  For apps with
encrypted models, the pipeline automatically generates the analysis scripts and
send them to the dynamic analyzer, \toolmxt, which performs a non-sophisticated
form of in-memory extraction of model representations. \toolmxt represents a
realistic attacker who attempts to steal the ML models from an app installed on
her own phone. Models extracted this way are in plaintext formats, even though
they exist in encrypted forms in the device storage or the app packages. Our
evaluation of \toolmxr and \toolmxt (\S\ref{evalmxr} and \S\ref{sec:modverify})
shows that they are highly accurate for our use, despite the simple analysis
techniques. We report our findings and insights drawn from the large-scale
analysis results produced by \toolmxr and \toolmxt in \S\ref{q1findings} and
\S\ref{sec:q2findings}, respectively. 

We investigated both the financial impact and the security impact of model leakages. 
For financial impact, we found that the attackers would benefit
from the savings of model licenses fee and Research \& Development (R\&D) investment; 
while the model vendors would suffer from losing pricing advantages and market share. 
The security impact includes easier bypass of model based access control and further
security and privacy breaches, which could affect both the end users and the model vendors.
(\S\ref{sec:financial}). 

%% file: static_new.tex
\section{Q1: \Qone} \label{sec:static}


\begin{figure*}[!h]
  \centering
      \includegraphics[width=0.7\textwidth]{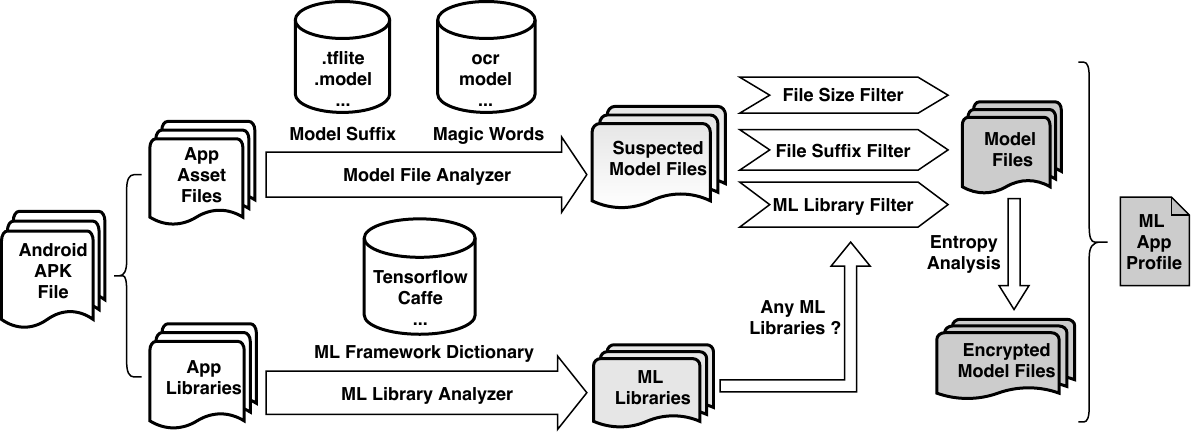}
      
              \caption{Identify Encrypted Models with \toolmxr}
              \medskip
              \small
                \toolmxr extracts an app's asset files and libraries from the APK file,
                analyzes the native libraries and asset files to identify ML frameworks, SDK libraries
                and model files. Then it applies model filters combining file sizes, file suffixes and ML libraries to reduce
                false positives and use entropy analysis to identify encrypted models. \par
      \label{fig:staticanalysis}
\end{figure*}

\subsection{Android App Collection}
We collect apps from three Android app markets: Google Play, Tencent My App, and
360 Mobile Assistant. They are the leading Android app stores in the US and China
\cite{appchina}. We download the apps labeled \textit{TRENDING} and \textit{NEW}
across all 55 categories from Google Play (12,711), and all recently updated
apps from Tencent My App (2,192) and 360 Mobile Assistant (31,850).


\subsection{Methodology of \toolmxr}
\label{sec:xraysdk}

\toolmxr statically detects if an app uses on-device ML and whether or not its
models are protected or encrypted. \toolmxr is simple by design and adopts a
best-effort detection strategy that errs on the side of soundness (\ie low false
positives), which is sufficient for our purpose of analyzing model leakage. 

We only consider encrypted models as protected in this study. We are
aware that some apps obfuscate the description text in the models. As we will
discuss in Section~\ref{sec:counter}, obfuscation
may make it harder for the attacker to understand the model, but does not prevent 
the attacker from reusing it at all.

The workflow of \toolmxr is shown in Figure \ref{fig:staticanalysis}. For a
given app, \toolmxr disassembles the APK file and extracts the app asset files
and the native libraries. Next, it identifies the ML libraries/frameworks and
the model files as follows:   

\point{ML Frameworks and SDK Libraries}
On-device model inference always use native ML libraries for performance
reasons. Inspired by Xu's work~\cite{Xu2019}, we use keyword searching in
binaries for identifying native ML libraries. \toolmxr supports a configurable
dictionary that maps keywords to corresponding ML frameworks, making it easy to
include new ML frameworks or evaluate the accuracy of keywords(listed in Appendix~\ref{tab:mlmagicwords}).
%
Further, \toolmxr supports generic keywords, such as ``NeuralNetwork'',``LSTM'',
``CNN'', and ``RNN'' to discover unpopular ML frameworks. However, these generic
keywords may cause false positives. We evaluate and verify the results in
\S\ref{evalmxr}.


\point{ML Model Files}
To identify model files, previous work \cite{Xu2019} rely on file suffix match
to find models that follow the common naming schemes. We find, however, many
model files are arbitrarily named. Therefore, We use a hybrid
approach combining file suffix match and path keyword match (\eg {\tt
../models/arbitrary.name} can be a model file). We address false positives by
using three filters: whether the file size is big enough (more than 8 KB);
whether it has a file suffix that is unlikely for ML models (\eg {\tt model.jpg});
whether the app has ML libraries.

\point{Encrypted Model Files}
We use the standard entropy test to infer if a model file is encrypted or not.
High entropy in a file is typically resulted from encryption or compression
\cite{entropywiki}. For compressed files, we rule them out by checking file
types and magic numbers. We use 7.99 as the entropy threshold for encryption in
the range of [0,8], which is the average entropy of the sampled encrypted model
files (see \S\ref{evalmxr}). Previous work\cite{Xu2019} treats models that
cannot be parsed by ML framework as encrypted models, which is not suitable in
our analysis and has high false positives for several reasons, such as the lack
of a proper parser, customized model formats, aggregated models, \etc

\point{ML App Profiles}
As the output, \toolmxr generates a profile for each app analyzed. A profile
comprises of two parts: ML models and SDK libraries. For ML models, it records
file names, sizes, MD5 hash and entropy. In particular, the MD5 hashes help us
identify shared/reused models among different apps (as discussed in
\S\ref{q1findings}).

For SDK libraries, we record framework names, the exported symbols, and the
strings extracted from the binaries. They contain information about the ML
functionalities, such as OCR, face detection, liveness detection. Our analysis
pipeline uses such information to generate the statistics on the use of ML
libraries (\S\ref{q1findings}).


\subsection{Accuracy Evaluation of \toolmxr} \label{evalmxr}
\point{Accuracy of Identifying ML Apps}
To establish the ground truth for this evaluation, we chose the 219 non-ML apps labeled
by \cite{Xu2019} as the true negatives, and 
we manually selected and 
verified 219 random ML apps as the true positives. 
We evaluated \toolmxr on this set of 438
apps. It achieved a false negative rate of 6.8\% (missed 30 ML apps) and a false
positive rate of 0\% (zero non-ML apps is classified as ML apps). We checked the
30 missed ML apps, and found out that they are using unpopular ML Frameworks
whose keywords are not in the dictionary. We found two ML apps that \toolmxr
correctly detected but are missed by \cite{Xu2019}, one using ULSFaceTracker,
which is an unpopular ML framework and the other using TensorFlow.

To further evaluate the false positive rate, we run \toolmxr on our entire set
of 46,753 apps and randomly sampled 100 apps labeled by \toolmxr as ML apps (50
apps from Google Play and 50 apps from Chinese app market). We then manually
checked these 100 apps and found 3 apps that are not ML apps (false positive
rate of 3\%). The manual check was done by examining the library's exposed
symbols and functions. This relatively low false positive rate shows \toolmxr's
high accuracy in detecting ML apps for our large-scale study.

\point{Accuracy of Identifying Models}
We randomly sampled 100 model files identified by \toolmxr from Chinese app
markets and Google Play, respectively, and manually verified the results.
\toolmxr achieved a true positive rate of 91\% and 97\%, respectively. 

In order to evaluate how widely apps conform to model standard naming
conventions, we manually checked 100 ML apps from both Google Play and Chinese
app market and found 24 apps that do not follow
any clear naming conventions. Some use "\textit{.tfl}" and "\textit{.lite}"
instead of the normal "\textit{.tflite}" for TensorFlow Lite models. Some use
"\textit{3\_class\_model}" without a suffix. Some have meaningful but not
standard suffixes such as "\textit{.rpnmodel}","\textit{.traineddata}". Other
have very generic suffixes such as "\textit{.bin}", "\textit{.dat}", and
"\textit{.bundle}". This observation shows that file suffix matching alone can
miss a lot of model files. Table~\ref{tab:model-suffix} shows the top 5 popular
model file suffixes used in different app markets. Many of these popular
suffixes are not standard. \toolmxr's model detection does not solely depend on
model file names.  

 \begin{table}[!ht]
  \caption{Popular model suffix among different app markets}
  \label{tab:model-suffix}
  \centering\footnotesize
  \begin{tabular}{c|c|c|c} \hline
  \begin{tabular}[c]{@{}c@{}}360 Mobile  \\ Assistant \end{tabular} &
  Num.Of.Cases&
  \begin{tabular}[c]{@{}c@{}}Google\\ Play \end{tabular} &
  Num.Of.Cases \\ \hline
      \textit{.bin} & 1860 & \textit{.bin} & 318 \\
      \textit{.model} & 1540 & \textit{.model} & 175 \\
      \textit{.rpnmodel} & 257 & \textit{.pb} & 93 \\
      \textit{.binary} & 212 & \textit{.tflite} & 83 \\
      \textit{.dat} & 201 & \textit{.traineddata} & 46 \\
      \hline
  \end{tabular}
  \end{table}

\begin{figure}[!h]
  \centering
      \includegraphics[width=0.4\textwidth]{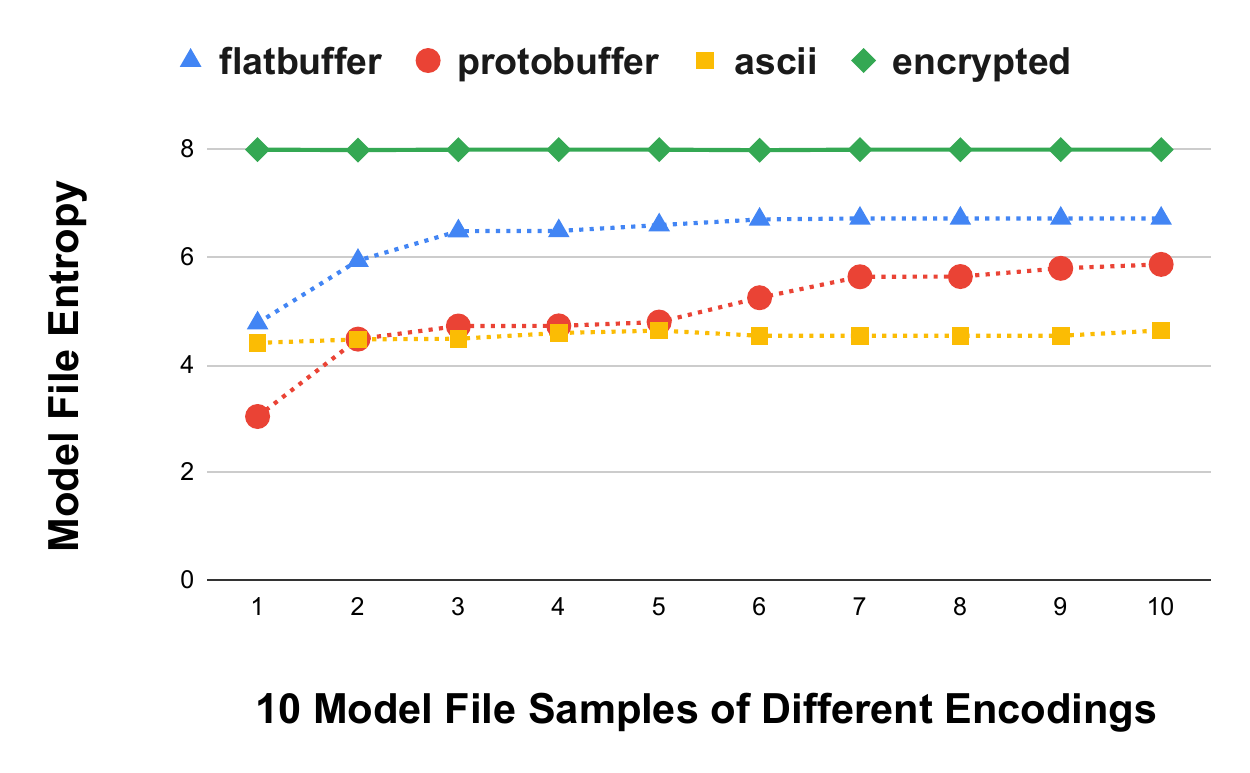}
      \caption{Model File Entropy of 4 Popular Encodings}
      \label{fig:entropy-range}
\end{figure}

\begin{figure}[!h]
  \centering
      \includegraphics[width=0.4\textwidth]{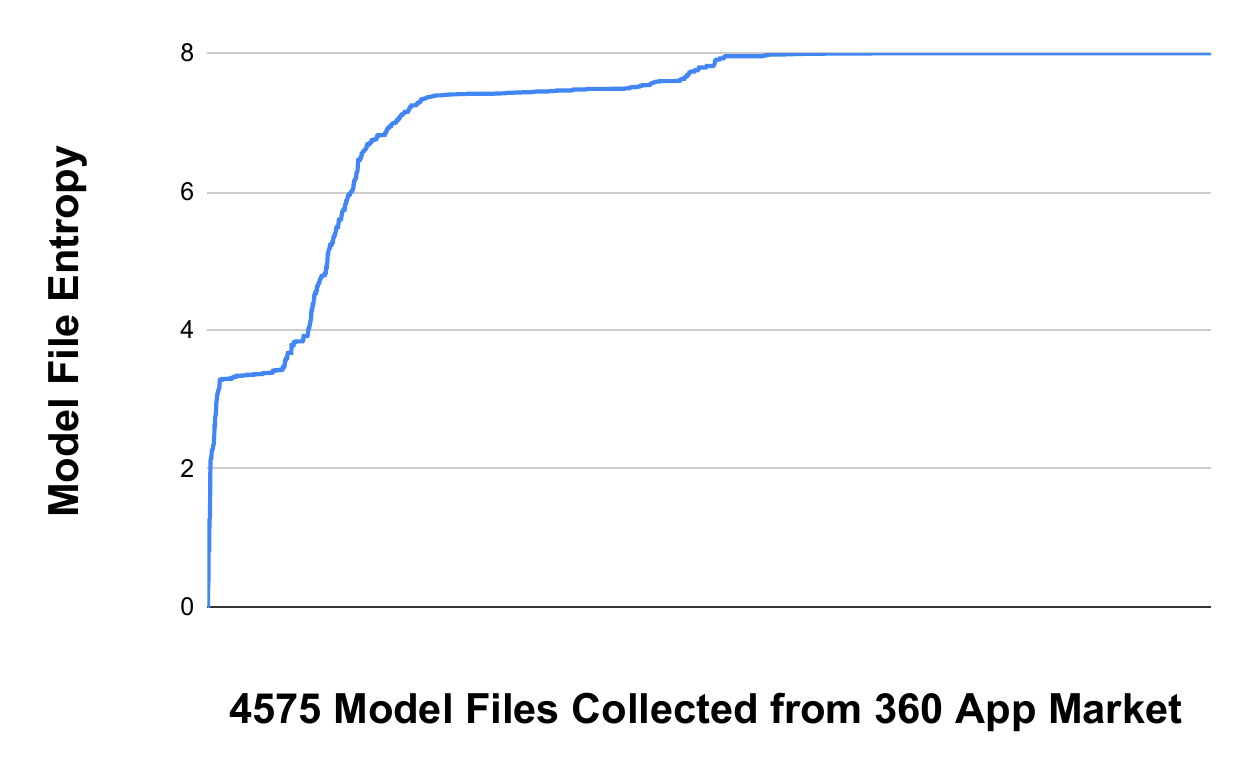}
      \caption{Model File Entropy Distribution of 360 App Market}
      \label{fig:entropy-dist}
\end{figure}

\point{Accuracy of Identifying Encrypted Models}
To evaluate whether entropy is a good indicator of encryption,
we sampled 40 models files from 4 popular encodings: ascii text,
protobuffer, flatbuffer, and encrypted format (10 for each category). 
As shown in Figure~\ref{fig:entropy-range}, the entropies of encrypted model files 
are all close to 8. The other encodings's entropies are significantly lower than 8. 
Figure \ref{fig:entropy-dist}
shows the entropy distribution of all model files collected from 360 App
Assistant app market. It shows that the typical entropy range of unencrypted model
files is between 3.5 and 7.5. 

\subsection{Findings and Insights} \label{q1findings}

We now present the results from our analysis as well as our findings and
insights, which provide answers to the question ``Q1: \qone''. We start with the
popularity and diversity of on-device ML among our collected apps, which echo
the importance of model security and protection. We then compare model
protection used in various apps. Especially, we draw observations on how model
protection varies across different app markets and different ML frameworks. We
also report our findings about the shared encrypted models used in different apps. 
In addition, we measured the adoption of GPU acceleration in ML apps and compared
the use of remote models and on-device models to further reveal the trends of on-device
model inference in mobile apps.

\point{Popularity and Diversity of ML Apps}
In total, we are able to collect 46,753 Android apps from Google Play, Tencent
My App and 360 Mobile Assistant stores. Using \toolmxr, we identify \mlappnum apps
that use on-device ML and have ML models deployed on devices, which accounts
for \mlapprate of our entire app collection. 

We also measure the popularity of ML apps for each category, 
as apps from certain categories may be more likely to use on-device ML
than others.  We used the app category information from the three app markets.
Table \ref{tab:appinfo1} shows the per-category numbers of total apps and ML
apps (i.e., apps using on-device ML). Our findings are summarized as follows:

\insight{On-device ML is gaining popularity in all categories.}  
There are more than 50 ML apps in each of the categories, which suggests the
widespread interests among app developers in using on-device ML. Among all the
categories, ``Business", ``Image" and ``News" are the top three that see most
ML apps. This observation confirms the diversity of apps that make heavy use of
on-device ML. It also highlights that a wide range of apps need to protect
their ML models and attackers have a wide selection of targets.  


\insight{More apps from Chinese markets are embracing on-device ML.}
This is reflected from both the percentage and the absolute number of ML apps:
Google Play has 178 (1.40\%), Tencent My App has 159 (7.25\%), and 360 Mobile
Assistant has 1,131 (3.55\%). 


As we can see from the above findings, Chinese app markets show a significant 
higher on-device machine learning adoption rate and unique property of per-category popularity,
making it a non-negligible dataset for studying on-device machine learning model protection.

\begin{table}[!ht]
\centering
\caption{The number of apps collected across markets.}
\label{tab:appinfo1}
\resizebox{\columnwidth}{!}{%
\begin{tabular}{r|r|r|r|r|r|r|r|r}
\hline
 & \multicolumn{2}{c|}{\begin{tabular}[c]{@{}c@{}}Google\\Play\end{tabular}} & \multicolumn{2}{c|}{\begin{tabular}[c]{@{}c@{}}Tencent\\ My App\end{tabular}} & 
 \multicolumn{2}{c|}{\begin{tabular}[c]{@{}c@{}}360\\ Mobile\\ Assistant\end{tabular}} &  \multicolumn{2}{c}{\begin{tabular}[c]{@{}c@{}}Total\end{tabular}} 
 \\ \hline
Category & All    & ML  & All   & ML & All   & ML  & All    & ML\\ \hline
\textbf{Business} & 404    & 2   & 99    & 2  & 2,450 & 296 & 2,953  & \textbf{300}\\ 
\textbf{News}     & 96     & 0   & 102   & 5  & 2,450 & 180 & 2,648  & \textbf{185}\\ 
\textbf{Images}   & 349    & 36  & 158   & 23 & 4,900 & 156 & 5,407  & \textbf{215}\\ 
Map      & 263    & 4   & 206   & 14 & 2,450 & 83  & 2,919  & 101\\ 
Social   & 438    & 23  & 141   & 17 & 2,450 & 79  & 3,029  & 119\\ 
Shopping & 183    & 5   & 112   & 16 & 2,450 & 84  & 2,745  & 105\\ 
Life     & 1,715  & 15  & 193   & 16 & 2,450 & 53  & 4,358  & 84 \\
Education& 389    & 3   & 116   &  7 & 2,450 & 74  & 2,955  & 84 \\
Finance  & 123    & 6   & 76    & 21 & 2,450 & 55  & 2,649  & 82 \\
Health   & 317    & 5   & 115   & 3  & 2,450 & 42  & 2,882  & 50 \\
Other    & 8,434  & 79  & 874   & 35 & 4,900 & 29  & 14,208 & 143\\ \hline
\textbf{Total} 
         & 12,711 & 178 & 2,192 & 159& 31,850&1,131&46,753  &1,468\\ \hline
\end{tabular}
}
\footnotesize
{\raggedright \textit{Note}: In 360 Mobile Assistant, the number of unique apps is 31,591 (smaller than 32,850) because some apps are multi-categorized. Image category contains 4,900 apps because we merged image and photo related apps.\par}
\end{table}


We measure the diversity of ML apps in terms of ML frameworks and
functionalities. We show the top-10 most common functionalities and their
distribution across different ML frameworks in Table \ref{tab:ml-frwk}. 

\insight{On-device ML offers highly diverse functionalities.}
Almost all common ML functionalities are now offered in the on-device fashion,
including OCR, face tracking, hand detection, speech recognition,
handwriting recognition, ID card recognition, and bank card recognition,
liveness detection, face recognition, iris recognition and so on. This high
diversity means that, from the model theft perspective, attackers can easily
find targets to steal ML models for any common functionalities.

\insight{Long tail in the distribution of ML frameworks used in apps.}
Besides the well-known frameworks such as TensorFlow, Caffe2/PyTorch, and
Parrots, many other ML frameworks are used for on-device ML, despite their
relatively low market share. For instance, as shown in Table~\ref{tab:ml-frwk},
Tencent NCNN \cite{Nihui}, Xiaomi Mace \cite{ConvertingCode}, Apache MXNet
\cite{ApacheLearning.}, and ULS from Utility Asset Store \cite{UnityMaking} are
used by a fraction of the apps that we collected. Each of them tends to cover
only a few ML functionalities. In addition, there could be other unpopular ML
frameworks that our analysis may have missed. This long tail in the distribution
of ML frameworks poses a challenge to model protection because frameworks use
different model formats, model loading/parsing routines, and model inference
pipelines.

\begin{table*}[!h]
\centering
\footnotesize
\caption{Number of apps using different ML Frameworks with different functionalities.}
\label{tab:ml-frwk}
\resizebox{0.9\textwidth}{!}{%
\begin{tabular}{r|r|r|r|r|r|r|r|r|r}
\hline
Functionality & \begin{tabular}[c]{@{}c@{}}TensorFlow \\ (Google)\end{tabular} & \begin{tabular}[c]{@{}c@{}}*Caffe2/PyTorch \\ (Facebook)\end{tabular} & \begin{tabular}[c]{@{}c@{}}*Parrots \\ (SenseTime)\end{tabular} & \begin{tabular}[c]{@{}c@{}}TFLite \\ (Google)\end{tabular} & \begin{tabular}[c]{@{}c@{}}NCNN \\ (Tencent)\end{tabular} & \begin{tabular}[c]{@{}c@{}}Mace \\ (Xiaomi)\end{tabular} & \begin{tabular}[c]{@{}c@{}}MxNet \\(Apache)\end{tabular} & \begin{tabular}[c]{@{}c@{}}ULS (Utility\\Asset Store)\end{tabular} & Total \\
\hline
 OCR(Optical Character Recognition)     & 41 & 186 & 140 & 6 & 37 & 18 & 1  & 11 & 441 \\
 Face Tracking              & 26 & 272 & 216 & 7 & 53 & 6  & 13 & 27 & 620 \\
 Speech Recognition      & 7  & 32  & 9   & 1 & 11 & 18 & 1  & 9  & 88 \\
 Hand Detection             & 4  & 0   & 0   & 2 & 4  & 0  & 0  & 0  & 10   \\ 
 Handwriting Recognition & 8  & 17  & 1   & 0 & 16 & 0  & 0  & 0  & 42 \\
 \textbf{Liveness Detection}        & 32 & 392 & 349 & 9 & 70 & 7 & 10 & 3 & 872 \\
 \textbf{Face Recognition} & 17 & 116 & 95  & 6 & 40 & 7 & 10 & 3 & 294 \\
 \textbf{Iris Recognition}            & 0  & 4   & 0   & 0 & 2  & 0 & 3  & 0 & 9 \\
 ID Card Recognition  & 26 & 230 & 147 & 5 & 47 & 18 & 0 & 10 & 483 \\
 Bank Card Recognition & 11 & 126 & 117 & 2 & 16 & 18 & 0 & 9  & 299 \\ 
\hline
\end{tabular}
}
\\
\scriptsize
{\textit{Note}: 1) One app may use multiple frameworks for different ML
functionalities. Therefore, the sum of apps using different functionalities is
bigger than the number of total apps. 2) Security critical functionalities are in \textbf{bold
fonts} and can be used for fraud detection or access control.  3)
\textit{*}Caffe was initially developed by Berkeley, based on which Facebook
built Caffe2, which was later merged with PyTorch. The following uses ``Caffe''
to represent Caffe, Caffe2 and PyTorch. 
}
\end{table*}

\input{modelupdate.tex}

\begin{figure*}[!h]
\centering
  \begin{subfigure}[b]{0.32\textwidth}
    \includegraphics[height=3.2cm]{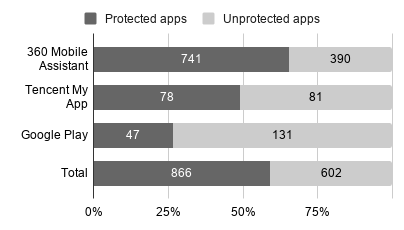}
     \caption{\centering Apps using protected/encrypted models vs. those using unprotected models}
     \label{fig:app-enc}
  \end{subfigure}
  \begin{subfigure}[b]{0.32\textwidth}
    \includegraphics[height=3.2cm]{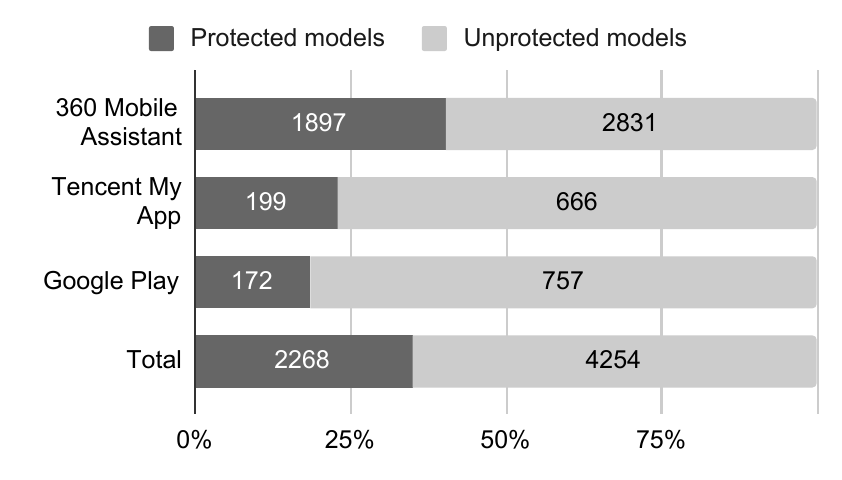}
     \caption{\centering On-device models that are protected/encrypted vs. those not}
     \label{fig:model-enc}
  \end{subfigure}
  \begin{subfigure}[b]{0.32\textwidth}
    \includegraphics[height=3.2cm]{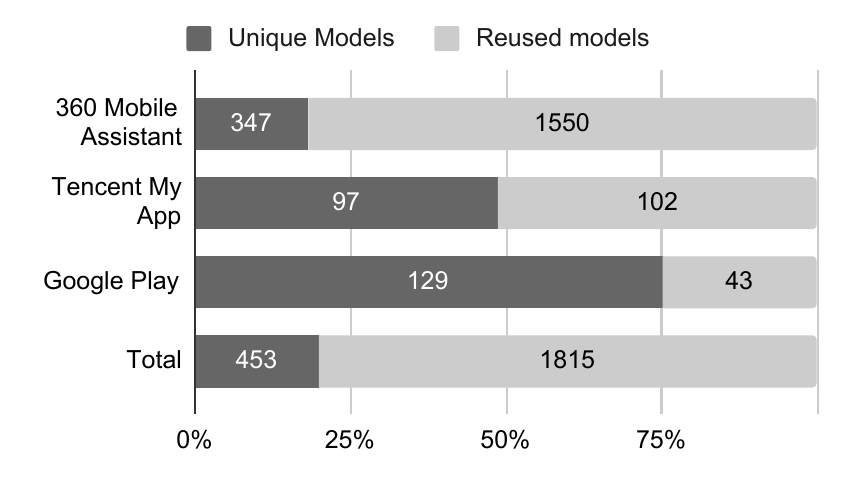}
    \caption{\centering Unique encrypted models vs. encrypted models reused/shared by multiple apps. }
    \label{fig:model-uniq}
  \end{subfigure}
  \caption{Statistics on ML model protection and reuse, grouped by app markets. The ``total'' number of unique models is less than the sum of the per-store numbers because some models are not unique from different stores.}
  \label{fig:model-security-store}
\end{figure*}

\point{Model Protection Across App Stores}
Figure \ref{fig:model-security-store} gives the per-app-market statistics on ML
model protection and reuse. Figure \ref{fig:app-enc} shows the per-market
numbers of protected apps (\ie apps using protected/encrypted models) and
unprotected apps (\ie apps using unprotected models). 

\insight{Overall, only 59\% of ML apps protect their models. }
The rest of the apps (602 in total) simply include the models in plaintext,
which can be easily extracted from the app packages or installation directories.
This result is alarming and suggests that a large number of app developers
are unaware of model theft risks and fail to protect their models. It also
shows that, for 41\% of the ML apps, stealing their models is as easy as
downloading and decompressing their app packages. We urge stakeholders and
security researchers to raise their awareness and understanding of model thefts,
which is a goal of this work.  


\insight{Percentages of protected models vary across app markets.}
When looking closer at each app market, it is obvious to see that Google Play
has the lowest percentage of ML apps using protected models (26\%) whereas 360
Mobile Assistant has the highest (66\%) and Tencent My App follows closely (59\%).
A similar conclusion can be drawn on the unique models (i.e., excluding reused models) 
found in those apps: 26\% models in Chinese apps are protected whereas the percentage 
of protected models in Google Play apps is 23\%. 
These percentages indicate that the apps from the Chinese markets are more
active in protecting their ML models, possibly due to better security awareness
or higher risks \cite{Face++Services,SenseTimest}.

When zooming into apps and focusing on individual models (\ie some apps use
multiple ML models for different functionalities), the percentages of unprotected
models (Figure \ref{fig:model-enc}) become even higher. Overall, 4,254 out of
6,522 models (77\%) are unprotected and thus easily extractable and reverse
engineered.



\begin{figure*}[!h]
\centering
  \begin{subfigure}[b]{0.32\textwidth}
    \includegraphics[width=\linewidth]{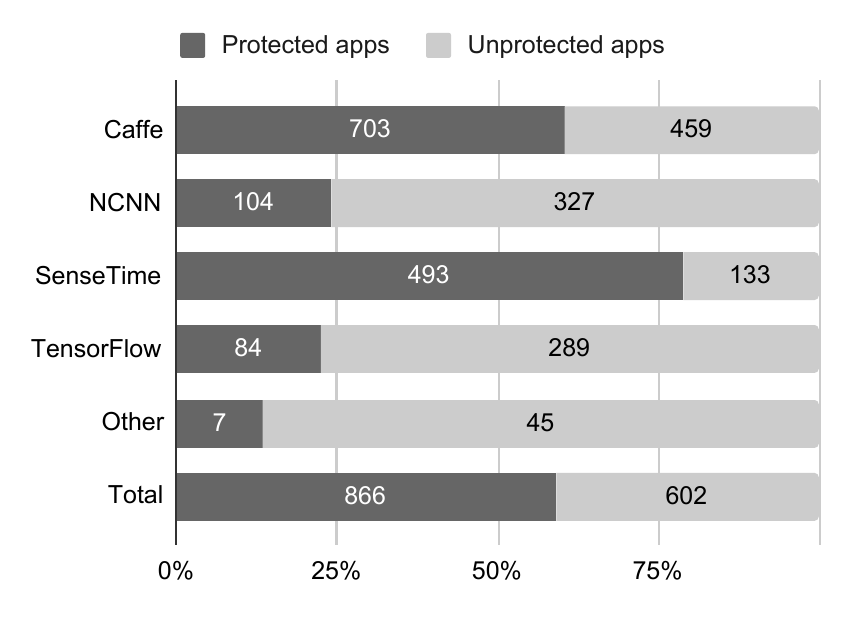}
    \caption{\centering Apps using protected models vs. those using unprotected models}
    \label{fig:app-fw-enc}
  \end{subfigure}
  \begin{subfigure}[b]{0.32\textwidth}
    \includegraphics[width=\linewidth]{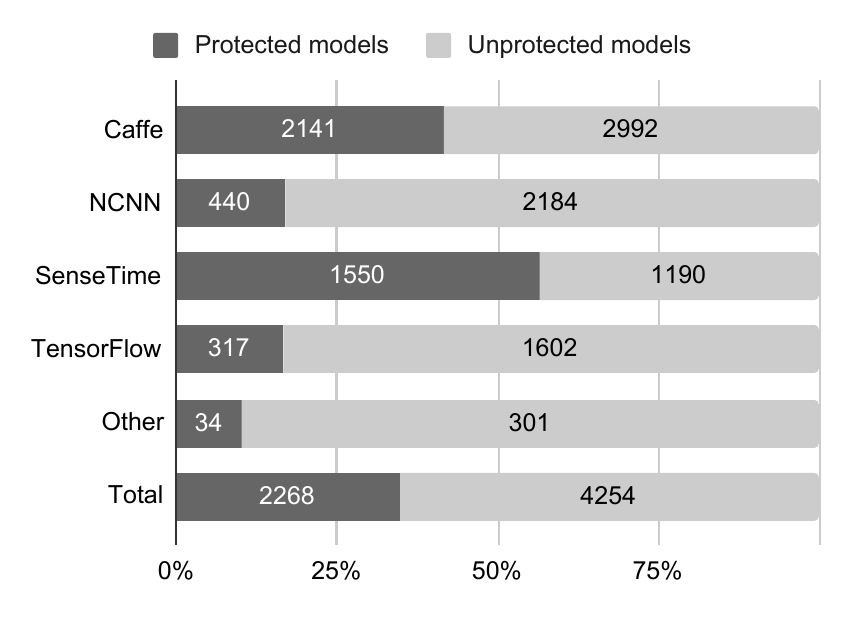}
    \caption{\centering On-device models that are protected/encrypted vs. those not}
    \label{fig:model-fw-enc}
  \end{subfigure}
  \begin{subfigure}[b]{0.32\textwidth}
    \includegraphics[width=\linewidth]{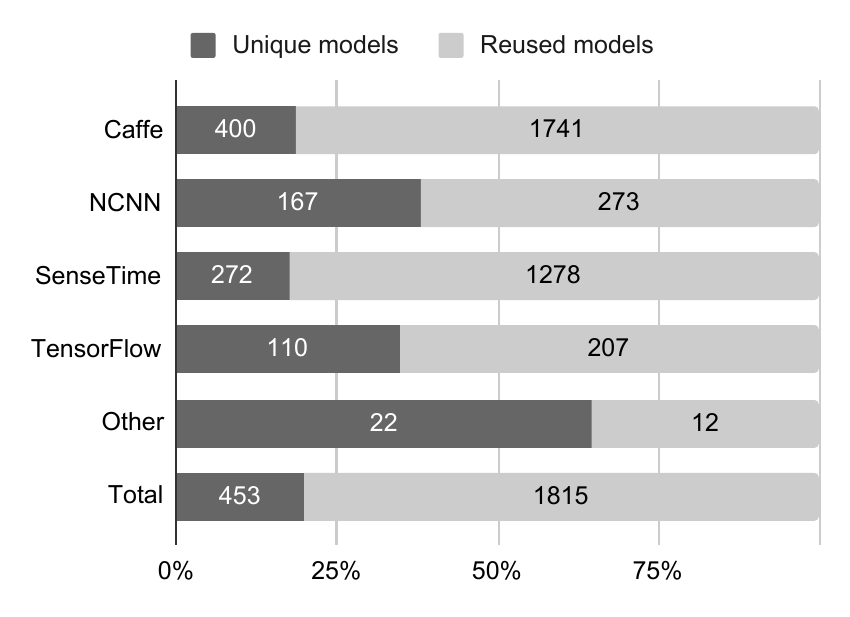}
    \caption{\centering Unique encrypted models vs. encrypted models reused/shared by multiple apps}
    \label{fig:model-fw-uniq}
  \end{subfigure}
  \caption{Statistics on ML model protection and reuse, grouped by ML frameworks. The ``total'' number is less than the sum of the per-framework numbers because many apps use multiple frameworks for different functionalities. }
  \label{fig:model-security-fw}

\end{figure*}

\point{Model Protection Across ML Frameworks}
We also derive the per-ML-framework statistics on model protection (Figure
\ref{fig:model-security-fw}). The frameworks used by a relatively small number
apps, including MXNet, Mace, TFLite, and ULS, are grouped into the ``Other"
category. 

\insight{Some popular ML frameworks have wider adoption of model protection,
but some not}. As shows in Figure \ref{fig:app-fw-enc}, more than
79\% of the apps using SenseTime (Parrots) have protected models, followed by
apps using Caffe (60\% of them have protected models). For apps using TensorFlow
and NCNN, the number is around 20\%. Apps using other frameworks are the least
protected against model thefts. 
This result can
be partly explained by the fact that some popular frameworks, such as SenseTime,
has first-party or third-party libraries that provide the model encryption
feature. However, even for apps using the top-4 ML frameworks, the percentage of
ML apps adopting model protection is still low at 59\%.   




\point{Encrypted Models Reused/Shared among Apps}
Our analysis also reveals a common practice used in developing on-device ML
apps, which has profound security implications. We found that many encrypted
models are reused or shared by different apps. The most widely shared model,
namely \textit{{\tt SenseID\_Motion\_Liveness.model}}, is found in 81 apps. 
This reuse might be legitimate given that app developers buy and use ML
models and services from third-party providers, such as SenseTime, instead of
developing their own ML features. 
The encrypted models reflect the awareness of the ML providers in preventing model thefts. 
However, we found 60 cases of different app companies are reusing model licenses.
One of the licenses is even used by 12 different app companies, indicating a high chance of illegal uses.



\insight{It is common to see the same encrypted model shared by different apps.}
For all the encrypted models that we detected from the apps, we calculate their
MD5 hashes and identify those models that are used in different and unrelated
apps. Figures \ref{fig:model-uniq} and \ref{fig:model-fw-uniq} show the numbers
of unique (or non-shared) models and reused (or shared) models, grouped by app
markets and ML frameworks, respectively. 
Overall, only 22\% of all the protected models are unique.
75\% of the
encrypted models from Google Play are unique whereas only 50\% and 19\% of the
encrypted models
on Tencent My App and 360 Mobile Assistant, respectively, are not
reused (Figure \ref{fig:model-uniq}). When grouped by ML frameworks, 82\% of
encrypted SenseTime models are shared, the highest among all frameworks (Figure
\ref{fig:model-fw-uniq}).  




\point{GPU Acceleration Adoption Rate among ML Apps}
Table ~\ref{tab:gpuusage} shows
the number ML apps and libraries that use GPU for acceleration. 
797(54\%) ML apps make use of GPU. \insight{The wide adoption
of GPU acceleration poses a challenge to the design of secure on-device ML.}
For instance, the naive idea of performing model inference and other model access operations entirely inside a trusted execution environment (TEE, \eg TrustZone) is not viable due to the need for GPU acceleration, which cannot be easily or efficiently accessed within the TEE. 

 \begin{table}[!ht]
  \caption{ML apps and libraries that use GPU acceleration}
  \label{tab:gpuusage}
  \centering\footnotesize
  \begin{tabular}{c|c|c|c} \hline
    \  &
  \begin{tabular}[c]{@{}c@{}}360 Mobile  \\ Assistant \end{tabular} &
  \begin{tabular}[c]{@{}c@{}}Tencent\\ My App \end{tabular} &
  \begin{tabular}[c]{@{}c@{}}Google\\ Play \end{tabular} \\ 
  \hline
     ML Apps       & 669 & 104 & 24 \\
  \hline
     ML Libraries  & 212 & 103 & 23 \\
      \hline
  \end{tabular}
  \end{table}

\input{remote.tex}

%% file: modelupdate.tex
\point{Models Downloaded at Runtime} \label{sec:model_update}
Mobile apps can always update on-device models as part of the app package update, or update models 
independently by downloading the models at runtime. After investigating a few open ML platforms 
including Android's Firebase and Apple's Core ML, we found that they support downloading models at runtime\cite{androidml, coreml}.
Other open-sourced ML platforms like Paddle-Lite~\cite{paddlelite}, NCNN~\cite{ncnn} and Mace~\cite{mace}, 
do not explicitly support downloading models at runtime. 
Developers who use their SDKs can implement this feature easily if they need it.
Some proprietary ML SDKs, like SenseTime, Face++, which are not open-sourced, do not leave enough information
for us to tell whether they implement this feature or not.

To measure how many ML apps that download models at runtime, we can use static analysis or dynamic analysis.
For dynamic analysis, we can run each app, monitor the downloaded files, and check
whether these files are ML models or not. It would require installing and running tens of thousands of apps, as well as 
triggering the model downloading process, which is not practical. For static analysis, we can reverse engineer 
each app and analyze whether it implements this feature or not. However, 
this feature can be implemented in a few lines of code without exporting any symbols and the app packages
are always obfuscated, making it hard to analyze.

We took an indirect approach. We measure the number of apps that contain on-device ML libraries but not 
any ML models. These apps have to download the models at runtime to use the ML function. We found 109 such apps, 
64 from the Chinese app markets and 45 from the US app markets.

%% file: remote.tex
\point{Measurement of Remote Models} \label{sec:remote}
Unlike on-device model inference, 
remote model inference allows 
an app to query a remote server with an object, and obtain the inference result from the response. 
Remote model inference does not necessarily leave footprints like 
machine learning libraries or models in the app packages. 
We thus measure the use of remote models through APIs provided by AI companies.

We investigated the APIs provided by notable AI companies from both US and China.
Given publicly available documentation, we were able to extract the use of remote models from Google Cloud AI, Amazon Cloud AI
and Baidu AI. Specifically, we scanned the API documentation for signature (unique naming) of remote ML inference libraries. 
For example, to use the remote Voice Synthesizer of Baidu AI, an app developer needs 
to include the library \textit{libBDSpeechDecoder\_V1.so}. 
We then collected all the signatures from the three companies, and analyzed the use of such signatures in our app collection.

We compared the number of apps using remote models, on-device models, or using both type of models in a hybrid mode. 
As Table~\ref{tab:remote} shows, 1,341 apps use remote models, 1,468 apps use on-device models, and 182 apps use both.
We emphasize again that on-device model inference is as popular as remote model inference.


 \begin{table}[!ht]
  \caption{Comparison between apps using remote and on-device ML models}
  \label{tab:remote}
  \centering\footnotesize
  \begin{tabular}{c|c|c|c|c} \hline
      App Number  &
  \begin{tabular}[c]{@{}c@{}}360 Mobile  \\ Assistant \end{tabular} &
  \begin{tabular}[c]{@{}c@{}}Tencent\\ My App \end{tabular} &
      \begin{tabular}[c]{@{}c@{}}Google\\ Play \end{tabular} & 
          Sum \\ 
  \hline
      Remote Models    &1,186 &   118& 37 & 1,341 \\
      On-device Models & 1,131 & 159& 178 & 1,468\\
      Hybrid Mode  & 153 & 23 & 6 & 182\\
      \hline
  \end{tabular}
  \end{table}

We also analyzed the type of ML services provided by remote models, and the coverage of remote models among Android apps.  
Among the 1,341 apps using remote models, 1,075 apps use NLP APIs (speech recognition/synthesizer, etc.), 
266 apps use ML Vision APIs (OCR, image labeling, landmark recognition, etc.). 
We did not find any security critical use cases for remote models. 
As we can see, remote ML models offer services such as NLP, Voice Synthesizer, OCR and so on, 
rather than liveness detection, face recognition, or other live image processing functionalities, as often seen in on-device models.
This indicates that on-device models are preferred in scenarios with security critical use cases, and real-time demands.
For the remaining scenarios, remote models are preferred for easier integration.

%% file: dynamic.tex
\section{Q2: \Qtwo} \label{sec:dynamic}
To answer this question, we build \toolmxt, a tool simple by design
to dynamically recover protected or encrypted models used in on-device
ML. Conceptually, \toolmxt represents a practical and unsophisticated attack, whereby an attacker installs apps on his or her own mobile device and uses the off-the-shelf app instrumentation tools to identify and export ML models loaded in the memory. 
\toolmxt mainly targets on-device ML models that are
encrypted during transportation and at rest (in storage) but not protected when in use or loaded in memory. 
For protected models mentioned in \S\ref{sec:static},
\toolmxt is performed to assess the robustness of
the protection.


\begin{figure*}[!h]
\centering
    \includegraphics[width=0.7\textwidth]{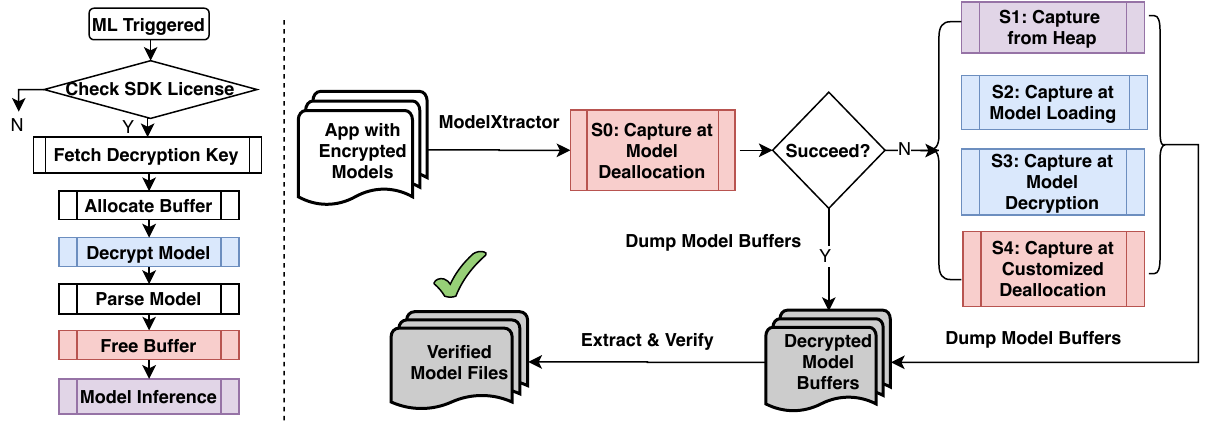}
     \caption{Extraction of (decrypted) models from app memory using \toolmxt 
     }
             \medskip
             \small
               The left side shows the typical workflow of model loading and decryption in mobile apps. The right side shows
               the workflow of \toolmxt. The same color on both sides indicate the same timing of the strategy being used. 
               The "Check SDK License" shows that a model provider will check an app's SDK license before releasing the decryption
               keys as a way to protect its IP.
     \label{fig:dynamicanalysis}
\end{figure*}

The workflow of \toolmxt is depicted in Figure \ref{fig:dynamicanalysis}. It
takes inputs from \toolmxr, including the information about the ML framework(s) and the model(s) used in the
app (described in \S\ref{sec:static}). 
These information helps to target and efficiently instrument an app during runtime,
and capture models in plaintext from the memory of the app. 
We discuss \toolmxt's code instrumentation strategies in \S\ref{sec:instr}, our
techniques for recognizing in-memory models in \S\ref{sec:modelrep}, and how
\toolmxt verifies captured models in \S\ref{sec:modverify}. Our findings,
insights, the answer to Q2, and several case studies are presented in
\S\ref{sec:q2findings} and \S\ref{sec:q2cases}. Responsible disclosure of our
findings is discussed in \S\ref{sec:disclosure}.



\subsection{App Instrumentation}
\label{sec:instr}

\toolmxt uses app instrumentation to dynamically find the memory buffers where
(decrypted) ML is loaded and accessed by the ML frameworks. 
For each app, \toolmxt determines which libraries and functions need to be
instrumented and when to start and stop each instrumentation, based on the
instrumentation strategies (discussed shortly). \toolmxt automatically generates
the code that needs to be inserted at different instrumentation points. It
employs the widely used Android instrumentation tool, Frida \cite{frida}, to
perform code injection. 

\toolmxt has a main instrumentation strategy (S0) and four alternative ones (S1-S4).  When the default strategy cannot capture the models, the alternatively strategies (S1-S4) will be used.  




\point{S0: Capture at Model Deallocation}
This is the default strategy since we observe the most
convenient time and place to capture an in-memory model is right before the
deallocation of the buffer where the model is loaded. This is because (1) memory
deallocation APIs (\eg {\tt free}) are limited in numbers and easy to
instrument, and (2) models are completely loaded and decrypted when their buffers
are to be freed. 

Naive instrumentation of deallocation APIs can lead to dramatic app slowdown. 
We optimize it by first only activating it after the ML library is loaded, and second, only for buffers greater than the minimum model size (a configurable threshold). To get buffer size, memory allocation APIs (\eg {\tt malloc}) are instrumented as well. The size information also helps correlate a decrypted model to its encrypted version (discussed in \S\ref{sec:modverify}).  

This default instrumentation
strategy may fail in the following uncommon scenarios. First, an app is not
using native ML libraries, but a JavaScript ML library.
Second, an app uses its own or customized memory allocator/deallocator. Third, a model
buffer is not freed during our dynamic analysis.





\point{S1: Capture from Heap}
This strategy dumps the entire heap region of an app when a ML functionality is
in use, in order to identify possible models in it. It is suitable for apps that
do not free model buffers timely or at all. It also helps in cases where
memory-managed ML libraries are used (\eg JavaScript) and buffer memory
deallocations (done by a garbage collector) are implicit or delayed. 

\point{S2: Capture at Model Loading}
This strategy instruments ML framework APIs that load models to buffers. We
manually collect a list of such APIs (\eg \textit{loadModel}) for the ML frameworks observed in our
analysis. This strategy is suitable for those apps where S0 fails and the ML
framework code is not obfuscated. 

\point{S3: Capture at Model Decryption}
This strategy instruments model decryption APIs  (\eg \textit{aes256\_decrypt}) in ML frameworks, which we 
collected manually. Similar to S2, it is not applicable to apps that use
obfuscated ML framework code.  

\point{S4: Capture at Customized Deallocation}
Some apps use customized memory deallocators. We manually identify a few such
allocators (\eg \textit{slab\_free}), which are instrumented similarly as S0.



\subsection{Model Representation and Recognition}
\label{sec:modelrep}

The app instrumentation described earlier captures memory buffers that may contain ML models. The next step is to perform model recognition from the buffers. The recognition is based on
the knowledge of in-memory model representations, \ie different ML
frameworks use different formats model encoding, discussed in the following.


Protobuf is the most popular model encoding format, used by TensorFlow, Caffe,
NCNN, and SenseTime. To detect and extract models in Protobuf from memory
buffers, \toolmxt uses two kinds of signatures: content signatures and encoding
signatures. The former is used to identify buffers that contain models and the
latter is used to locate the beginning of a model in a buffer.  

Model encoded in Protobuf usually contains words descriptive of neural
network structures and layers. For example, ``conv1" is used for one-dimension
convolution layer, and ``relu" for the Rectified Linear Unit. Such descriptive
words appear in almost every model and are used as the content signatures. 

The encoding signatures of Protobuf is derived from its encoding rule
\cite{protobufencoding}. For example, a Protobuf contains multiple {\em
messages}. Every message is a series of key-value pairs, or {\em fields}.
The key of a field is encoded as {\tt (field\_number $\ll$ 3) | wire\_type},
where the {\tt field\_number} is the ID of the field and {\tt wire\_type}
specifies the field type. 

A typical model in Protobuf starts with a message whose first field defines the
model name (\eg {\tt VGG\_CNN\_S}). This field usually has a {\tt wire\_type} of 2
(\ie a length-delimited string) and a {\tt field\_number} of 0 (\ie the first
field), which means that encoded key for this field is ``{\tt 0A}''. This key is usually
the first byte of a Protobuf encoded model. Due to
alignment, this key appears at a four-byte aligned address within the buffer. It
is used as an encoding signature.


Other model formats and representations have their own content and encoding
signature. For example, TFLite models usually include "TFL2" or “TFL3"
as version numbers. Some model files are even stored in JSON format, with easily
identifiable names for each field. 
%
%
Models from unknown frameworks or of unknown encoding formats are hard to
identify from memory. 
In such cases, we consider the buffer of the same size as the encrypted model
to contain the decrypted model. This buffer-model size matching
turns out to be fairly reliable in practice. The reason is that, when
implementing a decryption routine, programmers almost always allocate a buffer for
holding the decrypted content with the same size as the encrypted content. This
practice is both convenient (i.e., no need to precisely calculate the buffer
size before decryption) and safe (i.e., decrypted content is always shorter than
its encrypted counterpart due to the use of IV and padding during encryption). 
%
We show how buffer size matching is used in our case studies in
\S\ref{sec:q2cases}.

\subsection{Evaluation of \toolmxt}
\label{sec:modverify}
\point{Model Verification}
\toolmxt performs a two-step verification to remove falsely extracted models.
First, it confirms that the extracted model is valid. Second, it verifies that
the extracted model matches the encrypted model. We use publicly available model
parsers to verify the validity of extracted model buffers (\eg protobuf
decoder~\cite{pbDecoder} to extract protobuf content, and Netron~\cite{netron}
to show the model structure). When a decoding or parsing error happens, \toolmxt
considers the extracted model invalid and reports a failed model extraction
attempt. To confirm that an extracted model indeed corresponds to the encrypted
model, \toolmxt uses the buffer-model size matching described before.  

\point{Evaluation on Apps from Google Play} 
There are 47 ML apps from Google Play that use encryption to protect their
models. We applied \toolmxt on half of the ML apps (randomly selected 23 out of
47). 
Among the tested 23 apps, we successfully extracted decrypted models from 9 of
them. As for the other 14 apps, 2 apps do not use encryption, 1 app does not
using ML, and 11 apps do not have their models extracted for the following
reasons: apps cannot be instrumented;  apps
did not trigger the ML function; apps cannot be installed on our test devices.

\point{Evaluation on Apps from Chinese App Markets}
There are 819 apps from Chinese app markets found to be using encrypted models,
where model reuse is quite common as shown in our static analysis.
We carefully selected 59 of these apps 
prioritizing model popularity and app diversity. 
Our analyzed apps cover 15 of the top 45 most widely used models (\ie each is reused more than 10 times) 
and 8 app categories.

When analyzing the Chinese apps, we encountered some non-technical difficulties
of navigating the apps and triggering their ML functionalities. 
For instance, some apps require phone numbers from certain regions that we could
not obtain for user registration. A lot of them are online P2P loan apps or
banking apps that require a local bank account to trigger ML functionalities.
Out of the 59 apps, we managed to successfully navigate and trigger ML
functionalities in 16 apps. We then extracted decrypted models from 9 of them.

\point{Limitation of \toolmxt}
\toolmxt failed to extract 11 models whose ML functionalities were indeed
triggered. This was because of the limitation of our instrumentation strategies
discussed in \S\ref{sec:instr}. We note that these strategies and the design of
\toolmxt are not meant to extract every protected model. Instead, they represent
a fairly practical and simple attack,  designed only to reveal the insufficient
protection of ML models in today's mobile apps.

\subsection{Findings and Insights}
\label{sec:q2findings}



\point{Results of Dynamic Model Extraction}
Table \ref{tab:dump-model} shows the statistics on the 82 analyzed apps, grouped
by the ML frameworks they use. Among the 29 apps whose ML functionalities were
triggered, we successfully extracted models from 18 of them  (66\%). Considering
the reuse of those extracted encrypted models, the number of apps that are
affected by our model extraction is 347 (\ie 347 apps used the same models and
same protection techniques as the 18 apps that we extracted models from). This
extraction rate is alarming and shows that a majority of the apps using model
protection can still lose their valuable models to an unsophisticated attack. It
indicates that \insight{even for app developers and ML providers willing/trying
to protect their models, it is hard to do it in a robust way using the file
encryption-based techniques}. 

Table \ref{tab:model-cases} shows the per-app details about the extracted
models. We anonymized the apps for security concerns: many of them are highly
downloaded apps or provide security-critical services. Many of the listed apps
contain more than one ML models. For simplicity, we only list one representative
model for each app. 

\insight{Most decrypted models in memory are not protected at all}. As shown in
Table \ref{tab:model-cases}, most of the decrypted models (12 of 15) were easily
captured using the default strategy (S0) when model buffers are to be freed.
This means that the decrypted models may remain in memory for an extended period
of time (\ie decrypted models are not erased before memory deallocation), which
creates a large time window for model thefts for leakages. Moreover, this result
indicates that apps using encryption to protect models are not doing enough to
secure decrypted models loaded in memory, partly due to the lack practical
in-memory data protection techniques on mobile platforms.

\point{Popularity and Diversity of Extracted Models}
\insight{The extracted models are highly popular and diverse, some very valuable or security-critical.}
From Table \ref{tab:model-cases} we can see that 8 of 15 listed apps have been
downloaded more than 10 million times. Half of the extracted models belong to
commercial ML providers, such as SenseTime, and were purchased by the app
developers. Such models being leaked may cause direct financial loss to both app
developers and model owners (\S\ref{sec:financial}). 

As for diversity, the model size ranges from 160KB to 20MB. They span all
the popular frameworks, such as TensorFlow, TFLite, Caffe, SenseTime, Baidu, and
Face++. The observed model formats include Protobuf, FlatBuffer, JSON, and some
proprietary formats used by SenseTime, Face++ and Baidu. In terms of ML
functionalities, the models are used for face recognition, face tracking,
liveness detection, OCR, ID/card recognition, photo processing, and malware
detection. Among them, liveness detection, malware detection, and face
recognition are often used for security-critical purposes, such as access
control and fraud detection. Leakage of these models may give attackers an
advantage to develop model evasion techniques in a white-box fashion. 

\point{Reusability of the Extracted Models}
 Extracted models can be directly used by an attacker when they expect standard 
 input representations (e.g., images and video) and run on the common ML frameworks 
 (e.g., TensorFlow and PyTorch). More than 81\% of apps in our study contain directly 
 usable models. In some uncommon cases, such as the example given in Section~\ref{sec:q2cases},
 a model may expect a special/obfuscated input representation. Such a model, after extraction, 
 cannot be directly used. However, as we demonstrated in the paper, using standard reverse 
 engineering techniques, we could recover the feature vectors and reuse the extracted models in this case. 

\point{Potential Risk of Leaking SDK/Model License}
\insight{SDK/Model license are poorly protected.} Developers who bought the ML SDK license from model
provider usually ship the license along with app package. During analysis, we find the license are
used to verify legal use of SDK before model file get decrypted. However, license file are not protected
by the developer, which means it is possible to illegally use the SDK by stealing license file directly
from those apps that have bought it. Poor protection of license has been observed in both SenseTime
ML SDKs and some other SDKs, which actually affects hundreds of different apps.


 \begin{table}[!ht]
  \caption{Model extraction statistics.
  }
  \label{tab:dump-model}
  \centering\scriptsize
   \resizebox{\columnwidth}{!}{%
  \begin{tabular}{c|c|c|c|c|c} \hline
  \begin{tabular}[c]{@{}c@{}}ML \\ Framework \end{tabular} & 
  \begin{tabular}[c]{@{}c@{}}Unique Models\\Analyzed \end{tabular} &
  \begin{tabular}[c]{@{}c@{}}ML \\ Triggered\end{tabular} &
  \begin{tabular}[c]{@{}c@{}}Models \\ Extracted \end{tabular} &
  \begin{tabular}[c]{@{}c@{}}Models \\ Missed\end{tabular}  &
  \begin{tabular}[c]{@{}c@{}}Apps \\ Affected \end{tabular} \\ 
  \hline
  TensorFlow & 3 & 3 & 3 & 0 & 3 \\
  Caffe & 7 & 3 & 1 & 2 & 79\\
  SenseTime & 55 & 16 & 11 & 5 & 186 \\
  TFLite & 3 & 2 & 2 & 0 & 76\\
  NCNN & 9 & 3 & 0 & 3 & 0 \\
  Other & 5 & 3 & 2 & 1 & 88\\ \hline
  \textbf{Total} & 82 & 29 & 18 & 11 & 347 \\ \hline
  \end{tabular}
   }
\footnotesize
     \\ \textit{Note}: 347 is the sum of affected apps per framework after deduplication.
  \end{table}

\begin{table*}[!ht]
\centering
\footnotesize
\caption{Overview of Successfully Dumped Models with \toolmxt}
\label{tab:model-cases}
\resizebox{0.9\textwidth}{!}{%
\begin{tabular}{l|l|l|l|l|r|l|l} \hline
App name & Downloads & Framework & Model Functionality & Size (B) & Format & Reuses  & Extraction Strategy \\ \hline
\anony{com.basestonedata.instalment}{Anonymous App 1}  & 300M & TFLite    & Liveness Detection & 160K  & FlatBuffer & 18  & Freed Buffer \\
\anony{sweetsnap.lite.snapchat}{Anonymous App 2}        & 10M& Caffe      & Face Tracking         & 1.5M & Protobuf & 4  & Model Loading \\
\anony{com.camera.one.s10.camera}{Anonymous App 3}     & 27M & SenseTime & Face Tracking  & 2.3M & Protobuf & 77 &Freed Buffer  \\
\anony{com.dailyselfie.newlook.studio}{Anonymous App 4} & 100K& SenseTime & Face Filter    & 3.6M & Protobuf & 3   &Freed Buffer\\
\anony{video.like}{Anonymous App 5}                     & 100M& SenseTime  &Face Filter & 1.4M  & Protobuf  & 2 & Freed Buffer \\
\anony{io.anyline.example.store}{Anonymous App 6}       & 10K& TensorFlow & OCR  & 892K & Protobuf  &2  & Memory Dumping  \\
\anony{com.baseapp.eyeem}{Anonymous App 7}              & 10M& TensorFlow & Photo Process & 6.5M & Protobuf &1 & Freed Buffer  \\
\anony{com.decibel.beautycamera\_101}{Anonymous App 8}  & 10K& SenseTime & Face Track & 1.2M & Protobuf & 5  & Freed Buffer \\
\anony{com.hepai.hepaiandroid}{Anonymous App 9}         & 5.8M& Caffe & Face Detect & 60K & Protobuf & 77   & Freed Buffer  \\
\anony{com.lemon.faceu}{Anonymous App 10}                &10M& Face++ & Liveness & 468K & Unknown & 17    & Freed Buffer \\
\anony{com.campmobile.snowcamera}{Anonymous App 11}      & 100M& SenseTime & Face Detect & 1.7M & Protobuf & 18   & Freed Buffer \\
\anony{sino.cargocome.carrier.droid}{Anonymous App 12}   & 492K& Baidu  & Face Tracking  & 2.7M  & Unknown & 26  & Freed Buffer \\
\anony{com.taojinjia.charlotte}{Anonymous App 13}        & 250K& SenseTime  &  ID card & 1.3M  & Unknown & 13  & Freed Buffer  \\
\anony{com.kwai.video}{Anonymous App 14} & 100M& TFLite & Camera Filter & 228K & Json & 1  & Freed Buffer \\
\anony{com.anony.antivirus}{Anonymous App 15}  & 5K& TensorFlow  & Malware Classification & 20M & Protobuf &1 & Decryption Buffer  \\ \hline
\end{tabular}
}
\footnotesize
\\{\centering\textit{Note}: 1) We excluded some apps that dumped the same models as reported above; 2) We anonymized the name of the apps to protect the user's security; 3) Every app has
several models for different functionalities, we only list one representative model for each app.
\par}
\end{table*}

\subsection{Interesting Cases of Model Protection}
\label{sec:q2cases}
We observe a few cases clearly showing that some model providers use extra
protection on their models. Below we discuss these cases and share our insights.

\point{Encrypting Both Code and Model Files}
We analyzed an app that uses the Anyline OCR SDK. From the app profile generated by
ModelXRay, we can tell that this app uses TensorFlow framework.
It places the encrypted models under a directory named
``encrypted\_models''. Initially, ModelXtractor failed to extract the decrypted
models using the default strategy (S0). We manually investigated the reason and
found that, unlike most ML apps, this app runs ML inference in a customized
WebView, where an encrypted JavaScript, dynamically loaded at runtime, performs
the model decryption and inference. 
We analyzed the heap memory dumped by ModelXtractor using the alternative
strategy, S1, and found the TensorFlow model buffers
in the memory dump. We verified our findings by decoding the Protobuf model
buffers and extract the models' weights. 

It shows that, despite the extra protection and 
sophisticated obfuscation, the app can still lose its models to not-so-advanced
attacks that can locate and extract decrypted models in app memory.  

\point{Encrypting Feature Vectors and Formats}
When we analyzed one malware detection app, we found that it does not encrypt its model file.
Instead, it encrypts the feature vectors which is the input of the model. This app uses a Random Forest model
for malware classification. It uses TensorFlow framework and the model is in the format of Protobuf. 
There are more than one thousand features used in this malware
classification model, including the APIs used by the App, the Permissions claimed
in the Android Manifest files and so on. By encrypting the feature vectors, the developer
assumes it is impossible to (re)use the model because the input format and
content are unknown to attackers. However, we instrumented the decryption
functions and extracted the decrypted feature vectors. With this information, an
attacker can steal and recover the model as well as the feature vector format,
which can lead to model evasions or bypassing the malware detection.
It shows that even though some models take specific input format, with some basic
reverse engineering effort, the attacker can still uncover and reuse the model.

\point{Encrypting Models Multiple Times}
We also observed that one app encrypts
its models multiple times. This app offers online P2P
loans. It uses two models provided by SenseTime: one for ID card recognition and
the other for liveness detection, which are security critical. ModelXtractor successfully
extracted 6 model buffers, whose sizes range from 200KB to 800KB.
However, we only found 2 encrypted model files. When we were trying to map the
model buffers to the encrypted files, we found something very
interesting. 
One encrypted model file named
\textit{SenseID\_Ocr\_Idcard\_Mobile\_1.0.1.model} has a size of 1.3 MB. Among
the dumped model buffers, we have one buffer of the same size. It is supposed to
be the right decrypted buffer. After analyzing its
content, we found that it is actually a tar file containing multiple files, one
of which is \textit{align\_back.model}.
After inspecting the content of \textit{align\_back.model}, we found that it is
also an encrypted file. 
We then found another buffer of the same size, 246 KB, which contains a
decrypted model.
We finally realized that the app encrypts each model individually and 
compresses all encrypted models into a tar file, then encrypts it again.

\input{disclosure}

%% file: disclosure.tex
\subsection{Responsible Disclosure} \label{sec:disclosure}
We have contacted 12 major vendors whose apps have leaked models,
including Google, Facebook, Tencent, SenseTime and etc.
We have received responses from five of them.

In summary, for vendors that use plaintext models, one vendor is unaware of possible model leakage until we contact them.
For the other vendors, one of them is unaware of the impact that leaked models can incur.
Two vendors respond with lack of a practical solution to protect the models, 
in which one vendor is waiting for hardware support to encrypt the models securely, 
and the other fails to find an existing proprietary mitigation to make it harder for model reuse. 
This vendor assumes that malicious end users might eventually gain access to some model data, 
but not for practical use. 
For vendors whose models are encrypted but can still be extracted, 
our research raised internal discussions of one vendor on improving model security.
The vendor is taking actions on robust model protection, with research and collaborations with well-known security partners.

%% file: impact.tex
\section{Q3: \Qthree} \label{sec:financial}
ML models are the core intellectual properties of ML solution providers.
The impacts of leaked models are wide and profound, 
including substantial financial impact as well as significant security implications.

\subsection{Financial Impact}


\subsubsection{Financial Benefit for Attackers }
App developers usually have two legitimate ways to get ML models: 
(1) buying a license from ML solution providers, such as
SenseTime, Face++, and so on; (2) Developing their own
ML models, which usually requires a large amount of computing and human resources.
Stealing the models saves the attackers either the license fee paid to the model providers, 
or the research and development (R\&D) cost on the models.

\point{License Fee Savings for Attackers}
Usually, when vendors license an ML model, the app developer can choose between {\em online authorization} or {\em offline authorization}. 
A license with offline authorization allows a device to use the ML SDK without network connection. 
A company with such licenses is given unlimited uses on different devices \cite{license_fee}.
The down side is that the model provider has no control over the number of devices or which devices to have access to the model SDKs.
As a result, it is hard for the model provider to tell whether a model has been stolen or not.
According to Face++, the annual fee for a license with offline authorization is \$50,000 to \$200,000 \cite{license_fee}. 
The saving is large enough to motivate an attacker to steal the models or the model licenses.
In our analysis, we found 60 cases in which several different apps sharing one model license.
One of the licenses is even used by 12 different apps, indicating a high chance of illegal uses.

A license with online authorization can control the usages of the SDKs. 
Before using the model SDK, a device has to authenticate itself to the model provider with a license key. 
The model provider can then count the number of authorized devices, and charge the app company per device or per pack of devices.
Online authorization offers stronger protection of the model licenses than offline authorization. 
However, there are still chances that attackers stealthily use a license before it reaches the limit of the current pack. 
The market price for face landmark SDK is \$10,000 for up to 10,000 of online authorizations \cite{license_fee}.
Even though the savings are smaller than offline authorized licenses, attackers can still benefit from them financially.

\point{R\&D Savings for Attackers}
The R\&D cost of ML models comes from three sources: collecting and labeling
data for training, hiring AI engineers for designing and fine-tuning models, and
computing resources, such as renting or buying and maintaining storage servers and 
GPU clusters for training models. 

According to Amazon Mechanical Turk \cite{labelPrice}, the price of labeling an object ranges from \$0.012 to \$0.84, depending on the type of the object (e.g., image, text, semantic segmentation). 
Considering the CMU Multi-PIE database as an example, which contains more than 750,000 images \cite{cmupie}, the cost of labeling would be at least \$9,000. For larger databases, for example, MegaFace with 4.7 million labels \cite{megaface}, or some audio and video datasets \cite{audiodataset,videodataset}, the cost of labeling could be even higher.
According to LinkedIn statistics \cite{aiSalary}, the median base salary for machine learning engineers is \$145,000 per year.
Given a team with five engineers, training and fine-tuning a model for one year, the cost would be \$725,000.
Based on the pricing of Amazon SageMaker \cite{sageMakerPrice}, the monthly rate for ML storage is \$0.14 per GB, 
and the hourly rate for the current generation of ml.p3.2xlarge	accelerated computing is \$4.284. 
Still considering the CMU Multi-PIE database as an example, with a data size of 305GB, the yearly cost of 
data storage and training would be \$38,040.
 
Based on the above information, a conservative estimate on the total saving for attackers on model R\&D cost could be \$772,040.
Note that the salary of AI engineers are based on the public information of large AI companies, 
which can be higher than those from small companies. The number of AI engineers and the acutual model development cycle 
vary from case to case.
The estimation of R\&D cost should take all above factors into consideration.

\subsubsection{Financial Loss for Model Vendors  }
For vendors whose main business (source of income) depends on ML models, 
e.g., model providers or app companies,
model leakages result in pricing disadvantages, lost of customers and market share.

\point{Pricing Disadvantages for Vendors}
As mentioned earlier, the cost of ML models can reach millions of dollars, 
thereby competitors have strong motivation towards leaked models.
Once competitors start adopting leaked models with lower cost, they can offer lower prices to the customers.
At the same quality, customers are more willing to choose the cost efficient products. 
Therefore, vendors who leak their models will lose the pricing competition in the first place.

For model providers, the market is strongly competitive. 
In our study, we have found some top ML SDK providers, such as SenseTime, Megvii, Baidu, ULSee, Anyline, etc. 
Take Megvii as an example, according to Owler \cite{megviicompetitor}, 10 competitors are closely related to its businesses, such as Cognitec, SenseTime, Kairos, FaceFirst, Cortexica, etc.
For app companies, the competition is as much competitive if not more so. 
In Google Play only, our study found 36 apps using ML SDK for image recognition as the main business. 
Considering the other two stores, at least 215 apps are competing for this business.

\point{Anticipated Falling Market Share for Vendors}
The pricing disadvantage caused by leaked models will potentially result in loss of customers and market share, which will both lead to significant revenue loss. 
Take model provider SenseTime as an example, our study found 8 unique {\tt SenseID\_OCR} models, and each is reused by 21 apps on average. 
Loss of one single app customer will potentially bring a loss of at least \$10,000, based on the market price discussed earlier (e.g., \$10,000 for up to 10,000 of online authorizations).
In fact, SenseTime has more than 700 customers and partners \cite{sensetimecustomer}, and has a revenue of \$750 Million in 2019.
For app companies, we also observed unbalanced market share in the 215 apps competing for the business of image recognition. 
The number of downloads for these apps ranges from ten thousands to one hundred million. 
\insight{For both model providers and app companies, the decline in market share caused by pricing disadvantage may lead to further financial loss.}

\subsection{Security Impact}
Some ML models are used for security-critical purposes.
For example, liveness detection model is used to verify whether it is a real person holding a real ID card.
Face, fingerprint and iris recognition models are used to detect and verify the identity of a person.
These models bring in great convenience, for example, users do not need to go
to a bank or customer service centers to verify their identities.
However, breaches of such models bring in security and privacy concerns.

For attackers, a leaked security-critical model makes it
easier for them to design and craft adversarial examples. They
can then use the examples to either fake different identities,
or simply bypass the identity check of the apps \cite{fakefaces}.

We found more than 100 apps using on-device ML models for banking and loan services.
These apps provide personal loan services aiming at quick and convenient loan applications.
They use face recognition models to verify the identity of a person by taking a short video, and comparing with the photo on the ID card. The apps then determine the credit limits and rates to loan to the applicants.
When the models are leaked, attackers can easily fake identities of other applicants, and apply for loans on their behalf.


In our analysis, we found that 872 apps are using liveness detection
models, representing 59\% of all the apps using on-device ML.
We also found security-critical models to be shared among
different apps, for example, the {\tt SenseID\_Motion\_Liveness}
model is shared by 81 apps. Leakage of this 
model from any of the apps will make it easier for the attackers to bypass the detection 
to all the 81 apps.

For end users, it raises the concern that attackers with
faked identities can access users’ private information.
For example, some apps provide online medical services, such as booking appointments, 
filling out medical history forms, receiving electrical prescriptions, and laboratory reports from the doctors.
They may also use on-device ML models to verify the identities of patients.
Bypassing the verification will allow attackers to access personal medical records.
In our analysis, we found 6 such apps, which have been downloaded more than 9 million times on 360 Mobile Assistant Store. 
One of the face detection model, although encrypted, is shared by 77 different apps.
Leakage of the model from any of the apps will potentially expose the personal medical records of mass end users.
It is therefore important for vendors to protect the models, especially when they are security-critical.
\insight{Vendors and app developers should be careful about the potential security impact caused by leaked/stolen models. }

%% file: countermeasures.tex
\section{Countermeasures} \label{sec:counter}
In this section, we discuss several existing approaches to protecting on-device machine learning models 
and their limitations. We also share our insights in the future research of model protection.

\subsection{Current Model Protection}

\emphasize{Obfuscation} makes it harder for attackers to recover the model.
We observed that developers have implemented their own obfuscation/de-obfuscation mechanisms, which impose non-trivial programming overhead.
For example, NCNN can convert models into binaries where text is all striped, and Mace can convert a model to C++ code~\cite{ncnn,mace}.

\emphasize{Encryption} prevents the attackers from directly accessing the model from a downloaded APK. 
We observed that developers use encryption in many ways to protect their models, including the ML feature vectors, 
ML models, and the code to run model inferences. However, they all fall victim to our non-sophisticated dynamic analysis.


\emphasize{Customized model frameworks/formats} increase the effort for attackers to identify and reuse the models. 
We observed that customized or proprietary model formats, such as MessagePack (.model), pickle (.pkl), Thrift (.thrift), 
can be used to counter against model reverse engineering.
We also observed customized ML library running encrypted JavaScript in a customized WebView.

\subsection{Limitations}

\emphasize{Obfuscation} is vulnerable to devoted attackers who can recover the model with knowledge of binary decompilation.
Attackers can leverage program slicing and partial execution \cite{rasthofer2016harvesting,chen2017mass} to de-obfuscate Android apps \cite{bichsel2016statistical,wong2018tackling}, and further decompile and recover the obfuscated models.
Even without these knowledge, attackers can reuse the model as a black box.

\emphasize{Encryption} is vulnerable to attackers who can perform dynamic analysis and instrument app memory at runtime.
We have demonstrated it in Section \ref{sec:instr}.

\emphasize{Customized model frameworks/formats} are vulnerable to documentation leakage of the model frameworks/formats.
The documentation may come from internal attackers, or skilled and patient attackers who have good motivation to reverse engineer the model frameworks/formats.

\subsection{Future Works}
\emphasize{Secure hardware} is the most promising approach to protecting models on mobile devices. 
It has been demonstrated on desktop platforms. 
For example, recent advance in TF-Trusted \cite{tf-trusted} allows developers to run Tensorflow 
models inside of secure enclaves, such as Intel SGX \cite{intel-sgx}. 
Slalom \cite{tramer2018slalom} uses SGX during model inference, applies homomorphic encryption on each layer's input and outsources the computation of linear layers to GPU securely. Privado \cite{tople2018privado} uses SGX to mitigate side channel attacks of input inference. TensorScone \cite{kunkel2019tensorscone} also uses SGX to protect model inference but does not consider GPU. 
Graviton \cite{volos2018graviton} is proposed to make GPU a trusted execution environment with minimal hardware changes incurred. 
So far, research in this area focuses on cloud-end security. 

Future research should consider secure hardware backed model inference on mobile device. 
For example, Arm TrustZone \cite{armtrustzone} in mobile devices can be used to provide model protection. 
There are also some unique challenges that needs to be addressed on mobile devices.
Compared with desktop platforms, mobile devices are more restricted in computation resources, 
making it impractical to perform model inference entirely in TEE. 
Given the wide adoption of GPU on mobile devices, an effective model protection should also 
consider using the GPU for acceleration in a secure way. 

%% file: discussion.tex

\section{Discussion} \label{sec:discuss}

\input{manual_effort.tex}

\point{Research Insights}
\textit{White-box Adversarial Machine Learning.}
Previous research on adversarial machine learning has been focused on black-box threat models, 
assuming the model files are inaccessible. Our research shows that 
an attacker can easily extract the protected private models. 
As a result, more research on defending adversarial machine learning under white-box threat model is much needed
to improve the resiliency of those models used in security
critical applications. 

\textit{Model Plagiarism Detection.}
As machine learning models are not well protected, attackers, instead of training their own model, can steal their competitor’s model
and reuse it. As a result, model plagiarism detection is needed to prevent this type of attack. 
It is challenging because the attacker can retrain their model based on the stolen one, making it looks very different. 
We need research to detect model plagiarism and provide forensic tools for illegal model reuse analysis. 

\point{Limitations} \label{sec:limitation}
Since the goal of this paper is to show that even simple tools can extract on-device ML models in a large scale, \toolmxr and \toolmxt are limited by the straightforward design of keyword matching. We acknowledge that the scale of model extraction can be further improved by leveraging program slicing and partial execution \cite{rasthofer2016harvesting,chen2017mass}, and Android app de-obfuscation \cite{bichsel2016statistical,wong2018tackling}. Further, model encoding and content features are limited to well-known ML SDKs having documentation available, thereby we believe an extended knowledge base can further include special model encoding formats.

We note that our financial loss analysis is subjective and limited by the asymmetric information of R\&D cost and company revenue. The approach is used to emphasize the point that costs can be very high. A more comprehensive study can be carried out by stakeholders having real data of model leakage cases.


%% file: manual_effort.tex
\point{Manual analysis effort} \label{sec:manual}
Although \toolmxt can automatically generate instrumentation scripts
customized for the apps, manual effort is required in the dynamic analysis. 
As described in Section~\ref{sec:modverify}, 
some Chinese apps require registration with valid phone numbers or regional bank accounts before using ML models.
Manual effort is thus needed to feed in valid registration information.
To maximize the chance of triggering ML models, manual effort is also needed to 
fully navigate the apps with ML-related functionalities.
After the model is loaded and suspected model buffer dumped by \toolmxt, 
manual effort is needed to verify the start of the model based on the encoding signatures described in Section~\ref{sec:modelrep}.
Then we truncate the buffer and use a model decoder, e.g. protobuf, to parse the buffer and manually verify whether it is a ML model.

The amount of manual effort depends on how easy it is to trigger the ML functionality.
Some apps do not need registration and the ML models are loaded by default,
such as some AI camera apps, 
extracting their models takes less than an hour. 
In the worst cases, such as some P2P loan apps, 
whole ML models cannot be loaded without registration with valid phone numbers and regional bank accounts, 
it may take hours to extract the models. 
We therefore prioritize on apps whose models can be easily extracted, 
and budget 2 hours for each app among the 82 apps we analyzed in Table \ref{tab:dump-model}.

%% file: relatedwork_1.tex
\section{Related Work} \label{sec:related}

Motivated by hardware acceleration and efficiency improvement of deep neural networks \cite{lee2019ondevice}, on-device model inference becomes a new trend \cite{Xu2019}. This work empirically evaluates model security on mobile devices. It interacts with three lines of research: machine learning model extraction, adversarial machine learning, and proprietary model protection. 

To extract information from Android apps, prior works have used various techniques, such as memory instrumentation, 
program slicing and partial execution. 
For example, to detect Android malware, Hoffmann presents static analysis with program slicing on Smali code \cite{hoffmann2013slicing}.
DroidTrace \cite{zheng2014droidtrace} presents ptrace based dynamic analysis with forward execution capability. 
DroidTrace monitors selected system calls of the target process, and classifies the behaviors through the system call sequences.
Rasthofer combines program slicing and dynamic execution \cite{rasthofer2016harvesting} to further extract values from obfuscated samples, which include reflected function calls, sensitive values in native code, dynamically loaded code, and other anti-analysis techniques. 
Similar works include DeGuard \cite{bichsel2016statistical} and TIRO \cite{wong2018tackling}.
To extract the cryptographic key of a TLS connection, DroidKex \cite{taubmann2018droidkex} applies fast extraction of ephemeral data from the memory of a running process. It then performs partial reconstruction on the semantics of data structures.
ARTIST provides an Android Runtime Instrumentation Toolkit \cite{dresel2016artist}, which monitors the execution of Java and native code. 
ARTIST parses OAT executable files in memory to find classes and methods of interest, and locate internal structures of the Android Runtime.
AndroidSlicer combines asynchronous slicing for data modeling and control dependencies in the callbacks \cite{azim2019dynamic}. 
It can locate instructions responsible for model loading/unloading, and track responsible parts based on app inputs.
Similarly, CredMiner investigates the prevalent unsafe uses of developer credentials \cite{zhou2015harvesting}. 
It leverages data flow analysis to identify the raw form of the
embedded credential.
Our work also combines static and dynamic analysis on Android apps, however, with a different goal of machine learning model extraction.

Prior work on machine learning model extraction focuses on learning-based techniques targeting ML-as-a-service. 
Tramer et. al proposes stealing machine learning models via prediction APIs \cite{tramer2016stealing}, since ML-as-a-service may accept partial feature vectors as inputs and include confidence values with predictions. Then, Wang et. al \cite{wang2018stealing} extend the attacks by stealing hyperparameters. Other work includes stealing the functionality of the models \cite{orekondy2019knockoff,jagielski2019high}, querying the gradient to reconstruct the models~\cite{milli2018model}, exploratory attacks to reverse engineer the classifiers \cite{sethi2018data}, and side channel attacks to recover the models \cite{batina2018csi}. Our work is orthogonal to these study by targeting on-device model inference, assuming the attackers having physical access to the mobile devices running model inference.

Model extraction paves the road for adversarial machine learning.
Prior work \cite{huang2011adversarial,kurakin2016adversarial} fooling 
the models or bypassing the check is mostly under the black-box threat model. 
Once ML models become white-box, attackers can easily craft adversarial 
examples to deceive the learning systems. 
Our study shows white-box adversarial machine learning is a real threat to
on-device ML models.


To protect machine learning model as an intellectual property, watermark technique has been used to detect illegitimate model uses \cite{zhang2018protecting,adi2018turning}. Moreover, fingerprinting has been used to protect model integrity. 
Chen et al. encodes fingerprint \cite{chen2019deepattest} in DNN weights 
so that the models can be attested to make sure it is not tampered or modified. Our research supports it with the finding that model plagiarism is a realistic problem especially for mobile platforms.

%% file: conclusion.tex
\section{Conclusion} \label{sec:conclusion}
We carry out a large scale security analysis of machine learning model protection
on 46,753 Android apps from both the Chinese and the US app markets. 
Our analysis shows that on-device machine learning is gaining popularity in every
category of mobile apps, however, 41\% of them are not protecting
their models. For those are, many suffer from weak protection mechanisms, such as using the same 
encrypted model for multiple apps, and even the encrypted models can be easily recovered with our unsophisticated analysis. Our impact analysis shows that
model leakage can financially benefit attacks with as high as millions of dollars,  
and allow attackers to evade model-based authentication and access user private information.
Attackers both 
technically can and financially are motivated to steal models. We call for research into robust model protection. 



%% file: ack.tex
\section*{Acknowledgment}
The authors would like to thank the paper shepherd Prof. Konrad Rieck and the anonymous reviewers for their insightful
comments. This project was supported by
the National Science Foundation (Grant\#: CNS-1748334) and  
the Army Research Office (Grant\#: W911NF-18-1-0093). 
Any opinions, findings, and conclusions or recommendations expressed in this
paper are those of the authors and do not necessarily reflect the views of the
funding agencies.

%% file: appendix.tex
\begin{appendices}
\renewcommand\thetable{\thesection\arabic{table}}
\renewcommand\thefigure{\thesection\arabic{figure}}

\section{Keywords for ML Frameworks} 
    \label{sec:magicwords}
  \setcounter{figure}{0}
  \setcounter{table}{0}

\begin{table}[!ht]
\footnotesize
\centering
\caption{ML Framework Keywords}
\label{tab:mlmagicwords}
 \resizebox{\columnwidth}{!}{%
    \begin{tabular}{c|c|c|c}
    \hline
    \textbf{Framework} & \textbf{Magic Words} & \textbf{Framework} &\textbf{Magic Words}\\
    \hline
    TensorFlow & tensorflow & Caffe & caffe \\
    \hline
    MXnet & mxnet & NCNN & ncnn \\
    \hline
    Mace & libmace, mace\_input &SenseTime & sensetime, st\_mobile \\
    \hline
    ULS & ulstracker, ulsface & Other & neuralnetwork, lstm, cnn, rnn \\
    \hline
 \end{tabular}
}
\\
\footnotesize
{\raggedright \textit{Note}: ``TensorFlow Lite'' and ``TensorFlow'' are merged into one framework.
\par
}
\end{table}

\end{appendices}